\documentclass{article}
\usepackage[a4paper, margin=1in]{geometry} 
\usepackage[nodisplayskipstretch]{setspace}
\usepackage{amsthm,amsmath,amsfonts,amssymb}
\usepackage{subcaption}
\usepackage[shortlabels]{enumitem}
\usepackage{amssymb}
\usepackage{xcolor}
\usepackage{graphicx}
\usepackage{calc}
\usepackage{float}
\usepackage{tikz}
\usepackage{graphics} 
\usepackage{amsfonts}
\usepackage{algpseudocode}
\usepackage{verbatim}
\usepackage{adjustbox}
\usepackage{blindtext}
\usepackage{algorithm}
\usepackage{longtable}
\usepackage{setspace}
\usepackage{natbib}
\newcommand{\calD}{\mathcal D}
\newcommand{\calV}{\mathcal V}
\newcommand{\calG}{\mathcal G}
\newcommand{\given}{\;|\;}
\newcommand{\ind}{\overset{ind}{\sim}}
\newcommand{\iid}{\overset{i.i.d}{\sim}}

\def\*#1{\mathbf{#1}}
\DeclareMathOperator{\Var}{Var}
\DeclareMathOperator{\Cov}{Cov}
\DeclareMathOperator{\argmin}{argmin}

\newtheorem{proposition}{Proposition}[section]
\def\bSig\mathbf{\Sigma}

\newcommand{\blue}[1]{{\leavevmode\color{black}{#1}}}

\usepackage{longtable}
\usepackage{setspace}
\newcommand{\calU}{\mathcal U}
\def\*#1{\mathbf{#1}}

\usepackage{xr}
\usepackage{authblk}
\usepackage{abstract}
\title{Visibility graph-based covariance functions for scalable spatial analysis in non-convex \blue{partially Euclidean} domains}

\author[1]{Brian Gilbert\thanks{brian.gilbert@nyulangone.org}}
\author[2]{Abhirup Datta}
\affil[1]{Department of Population Health, NYU Grossman School of Medicine, New York, New York, U.S.A.}
\affil[2]{Department of Biostatistics, Johns Hopkins Bloomberg School of Public Health, Baltimore, Maryland, U.S.A.}

\begin{document}

\maketitle

\label{firstpage}

\begin{abstract}
We present a new method for constructing valid covariance functions of Gaussian processes for spatial analysis in irregular, non-convex domains such as bodies of water. Standard covariance functions based on geodesic distances are not guaranteed to be positive definite on such domains, while existing non-Euclidean approaches fail to respect the partially Euclidean nature of these domains where the geodesic distance agrees with the Euclidean distances for some pairs of points. Using a visibility graph on the domain, we propose a class of covariance functions that preserve Euclidean-based covariances between points that are connected in the domain while incorporating the non-convex geometry of the domain via conditional independence relationships. We show that the proposed method preserves the partially Euclidean nature of the intrinsic geometry on the domain while maintaining validity (positive definiteness) and marginal stationarity of the covariance function over the entire parameter space, properties which are not always fulfilled by existing approaches to construct covariance functions on non-convex domains. We provide useful approximations to improve computational efficiency, resulting in a scalable algorithm. We compare the performance of our method with those of competing state-of-the-art methods using simulation studies on synthetic non-convex domains. The method is applied to data regarding acidity levels in the Chesapeake Bay, showing its potential for ecological monitoring in real-world spatial applications on irregular domains.\\\end{abstract}

\section{Introduction}
 Much of spatial analysis concerns data collected over domains that are convex and Euclidean (e.g., an agricultural plot of land) or where the effect of irregular boundaries or non-convexity in parts of the domain can be ignored due to the scale of the analysis. \blue{However,} in some scientific contexts, the irregularity and concavity of the spatial domain cannot be ignored. \blue{
As an example, the Chesapeake Bay, the largest estuary in the United States with its numerous tributaries and inlets, presents a highly complicated non-convex geometry characterized by an irregular and fragmented coastline that significantly deviates from a simple, convex shape. The Chesapeake Bay is continually measured for chemical concentration levels throughout its extent to monitor its ecological health, and spatial-statistical methods aimed at extending the information contained in these data to unsampled locations need to acknowledge the geometry of the bay, as drastic differences in water quality in the various tidal tributaries of the bay that are close in Euclidean distance are well documented \citep{najjar2020alkalinity}.} %
 In such irregularly shaped bodies of water like bays, lakes, and estuaries, it is more appropriate to use the ``as the fish swims" distance (the length of the shortest path through the water) rather than the ``as the crow flies" distance (the length of the shortest path on Earth's surface), as noted by \citet{little} and \citet{rathbun1998spatial}. 
 
 Mathematically speaking, in a non-convex domain, the \textit{geodesic} distances with respect to the \textit{intrinsic} Euclidean metric do not match the ordinary Euclidean metric. However, using non-Euclidean distance measures with common covariance functions like the Mat\'{e}rn family may not always yield positive definite covariance matrices \citep{curriero}. Therefore, it is desirable to construct covariance functions based on distance measures that respect physical geometry while maintaining statistical validity.

There are many methods for spatial analysis in non-convex domains. These include the general-purpose multidimensional scaling \citep[MDS,][]{cox} that finds an embedding of the data locations into a Euclidean space by minimizing distortion of the geodesic distances in the original space \blue{and using Gaussian process (GP) models on the warped space,} best positive-definite 
approximation of the geodesic-distance based \blue{GP} covariance matrix \citep{davis}, stochastic differential equation based approaches \citep{niu,borovitskiy,bakka}, graph-Laplacian Gaussian process \citep[GLGP][]{dunson} that uses a weighted complete graph on the domain with locally Euclidean weights, and BORA-GP \citep{jin2022spatial} that uses nearest-neighbor Gaussian processes  \citep[NNGP,][]{datta2016,finley2019efficient} with neighbor sets conforming to the geometry of the domain. A review of these approaches is provided in Section \ref{web_app_A}.

We focus on spatial analysis on irregular domains like water bodies which are non-convex subsets of Euclidean spaces and are {\em partially Euclidean}, i.e., the geodesic distance exactly equals the Euclidean distance for any pair of points that are {\em connected in the domain}, i.e., the straight line connecting the two points lie fully in the domain. It is desirable for covariance functions on such domains to preserve this partially Euclidean nature of the distance metric, because a spatial analysis using the Euclidean metric would presumably be employed were the data collection restricted to a convex subset of this non-convex domain. To our knowledge, none of the aforementioned approaches possess this property. 

\blue{We propose a general approach to construct valid covariance functions on any irregular non-convex partially-Euclidean domain. 
Our proposed method, {\em visGP}, is based on creating a visibility graph between points in the domain that incorporates the structure of the geometry and barriers (which the method requires as known input, as opposed to GLGP), and subsequently applying the graphical method of ``covariance selection" as described in \cite{dempster} to obtain the desired covariance matrix.} The finite-dimensional covariance matrix is then extended to a valid positive definite covariance function on the entire non-convex domain. Theoretically, we show that \blue{visGP} possesses several desirable qualities. It preserves the partially Euclidean nature of the domain while respecting the irregular geometry. Formally, we show that, starting from a covariance function $C$ which is valid on the Euclidean domain $\mathbb R^d$, it is possible to derive a visGP covariance function $C^*$ for a non-convex \blue{partially Euclidean} subset $\calD\, \blue{\subset}\, \mathbb R^d$ such that $C^*$ leaves unchanged the covariances from $C$ among pairs of points \blue{in $\calD$} whose geodesic distance coincides with Euclidean distances; that is, those pairs of points which are connected %
in the domain, while satisfying a conditional independence (Markov) property for points not connected in the domain \blue{thereby respecting the non-convexity.} We also preserve marginal stationarity of the covariance function in the entire domain. These properties ensure that the analysis using our covariance function restricted to any convex subset of the non-convex domain would agree exactly with standard spatial analysis using Euclidean distances. Additionally, the Markov property yields covariances that exactly agree with the geodesic covariances on certain special domains. 

Our construction has some connections to the recent notable contribution BORA-GP \citep{jin2022spatial}, which uses directed graphs based on only Euclidean nearest neighbors. However, there are important differences between the approaches. 
BORA-GP requires an ordering of the locations, which leads to a lack of stationarity for highly irregular domains, while we use undirected graphs and exactly preserve marginal stationarity. Also, BORA-GP does not attempt to preserve any covariance values relative to the Euclidean model, even though the domain \blue{is partially Euclidean. We illustrate these differences in Section \ref{sec:simcov}}

We outline the construction and mathematical theory of visGP below. We then propose some pragmatic approximations for implementation with computational tractability. These include a chordal graph completion that yields closed-form likelihood for visGP in terms of the original Euclidean GP covariance function, and a novel graph subsampling approach that leverages the additivity of log-likelihoods over chordal graphs. We demonstrate the method and compare it to alternatives through simulations and analysis of Chesapeake Bay data. 

\section{Valid covariance functions on non-convex \blue{partially Euclidean} domains}\label{sec:valid}
\subsection{Finite-dimensional construction}\label{sec:finite}

We first present a general approach to construct valid spatial covariance matrices on any arbitrary finite set of locations in a non-convex \blue{partially Euclidean} domain. Subsequently, we will extend to valid covariance functions of Gaussian processes over the entire domain. 
\blue{Consider locations $s_1,...,s_n$ which are points in a non-convex partially Euclidean domain $\mathcal{D} \subset \mathbb{R}^d$. Define the adjacency matrix $A$ where $a_{ij}$ is 1 if $s_i$ and $s_j$ are {\em connected in the domain}, i.e., the line segment between $s_i$ and $s_j$ remains wholly within $\mathcal{D}$, and $a_{ij}$ is $0$ otherwise. Let $\mathcal{G} = (\mathcal{V},E)$ be the corresponding formal undirected graph, i.e, $\mathcal{V}=\{s_1, ..., s_n\}$ and $E=\{(i,j) \given a_{ij}=1\}$. %
In the geometry and artificial intelligence literature, such a graph is often called the {\em visibility graph} as, if $s_i$ and $s_j$ are not connected in the domain, there exists a boundary or barrier that prohibits seeing $s_i$ from $s_j$ and vice versa.}  

We start with any \blue{isotropic ({\em Euclidean})} covariance function $C$ that is valid (positive definite) on the Euclidean domain $\mathbb R^d$ and \blue{depends only on the Euclidean distances between points,} like %
the Mat\'ern, exponential, and Gaussian covariances.  %
Our construction will be agnostic to the specific choice of $C$. Spatial analysis within the non-convex domain $\calD$, using $C$ with Euclidean distances, is a valid but inappropriate choice as it ignores the geometry of the domain \citep{little,rathbun1998spatial}. Using $C$ but with the geodesic distances, although it seems reasonable, does not guarantee positive-definiteness \citep{curriero}.   

For many \blue{pairs of} points in $\calD$, the geodesic distances \blue{exactly} coincide with the Euclidean distance. In fact, the covariance function $C$ would be a perfectly valid choice for analyzing data within any convex subset of $\calD$. Hence, we desire a covariance function that both respects the irregular and non-convex geometry of the domain $\calD$, but also acknowledges this \textit{partially Euclidean} nature of the domain. Formally, given $C$ and the finite set of locations $\mathcal V=\{s_1,\ldots,s_n\}$, we seek a covariance function
$C^*$ with the following properties, letting $L=C^*(\calV,\calV)$ denote the covariance matrix induced by $C^*$ on $\calV$:
\begin{equation}\label{eq:covsel}
\begin{split}
    \blue{L_{ij} =C(s_i,s_j) \mbox{ for all } i = j \mbox{ or } (i,j) \in E,}\\
    (L^{-1})_{ij} = 0 \mbox{ for all } i \neq j \mbox{ such that } (i,j) \notin E. 
    \end{split}
\end{equation}

\blue{As $C$ is isotropic, $C(s,s)$ does not depend on $s$.}  %
\blue{So, if $L_{ii}=C(s_i,s_i)$ as posited in (\ref{eq:covsel}),} the new covariance function $C^*$ will \blue{be {\em marginally stationary} on $\calV$, i.e., if $w(\calV) \sim N(0,L)$ then $w(s_i) \overset{d}{=} w(s_j)$ for all $s_i,s_j \in \calV$.} The condition \blue{$L_{ij}=C(s_i,s_j)$ for $(i,j) \in E$} recognizes the partially Euclidean nature of the domain, imposing that the covariance of points connected in the domain is given by a standard Euclidean covariance function. This formalizes the belief that the original covariance function is suitable for through-domain distances since, for these connections, there is no interference by boundaries. Finally, the \blue{second} condition \blue{in (\ref{eq:covsel})} posits that two points that are not connected in the domain are conditionally independent, given all other observations. This is reasonable if the boundaries are seen as an impediment to correlation between the points that they separate. We show in Section \ref{sec:theory} how this Markov property leads to covariances agreeing with the geodesic covariances on certain domains. 

\cite{dempster}, a seminal work on \textit{covariance selection}, showed that given any positive definite matrix $K$ and a graph $\mathcal G=(\calV,E)$ with nodes indexed on the rows of $K$, there exists a unique positive definite matrix $L$ such that $L_{ij} = K_{ij}$ if $i=j$ or $(i,j) \in E$, and $(L^{-1})_{ij}=0$ if $(i,j) \notin E$. \cite{speed} gives an efficient \textit{iterative proportional scaling} algorithm to obtain $L$ given $K$ and the graph $\calG$. We denote such an $L$ derived from $K$ and $\calG$ using covariance selection as $L=CovSel(K,\calG)$. Hence, that a unique matrix $L=CovSel(C(\calV,\calV),\calG)$ exists satisfying all properties in (\ref{eq:covsel}) follows directly from Dempster's covariance selection using the positive definite matrix $K=C(\calV,\calV)$ and letting $\mathcal G$ be the visibility graph on the domain. 
We then specify a Gaussian process on $\calV$ simply  as 
\begin{equation}\label{eq:finite}
    w(\calV) \sim N(0,L), \mbox{ with } L=CovSel(C(\calV,\calV),\calG).
\end{equation} 
which satisfies $w(s_i) \overset{d}{=} w(s_j)$ (marginally stationary), $Cov(w(s_i),w(s_j)) = C(s_i,s_j)$ if $s_i$ and $s_j$ are connected in the domain (partially Euclidean), and $Cov(w(s_i),w(s_j) \given w(\calV) \setminus \{w(s_i),w(s_j)\}) = 0$ (Markov on points not connected in the domain).

\subsection{Process formulation}\label{sec:process}

The formulation in the previous section only presents a process (or its covariance function) restricted to an arbitrary but finite set of locations \blue{$\calV$, which we now extend to a valid Gaussian process over the entire domain $\calD$, while retaining the essential characteristics (marginal stationarity, partially Euclidean, and Markov).} %
For any location $s$ outside $\calV$, we find a \textit{neighbor set} $N(s)$ of up to $k$ locations in $\calV$ that are closest to $s$, %
while 
 enforcing two conditions. First, each location in $N(s)$ is connected in the domain to $s$. This ensures that we are not including a location in $N(s)$ that is close to $s$ in Euclidean distance but far away in the geodesic distance, as that would distort the geometry of the domain. We also require that the sub-graph of $\calG$ restricted to $N(s)$ is complete, i.e., all pairwise locations in $N(s)$ are connected in the domain to each other. 
This implies that the covariances among $N(s)$ are Euclidean, and, in turn, ensures that the resulting process has desirable properties as discussed in \blue{Propositions \ref{prop:marg} - \ref{prop:convex_union}.} %
We specify the conditional distribution for any $s \notin \calV$ as 
\begin{align}\label{eq:process}
    w(s) \given w(\calV) &\sim N(B(s)w(N(s)),F(s)), \mbox{ where } \nonumber \\
    B(s)&=C(s,N(s)) C(N(s),N(s))^{-1}, \mbox{ and }\\
    F(s) &= C(s,s)- C(s,N(s)) 
    C(N(s),N(s))^{-1}C(N(s),s) \nonumber.
\end{align}

Equations (\ref{eq:finite}) and (\ref{eq:process}) complete the specification of a Gaussian process $w(\cdot)$ on the entire domain $\calD$. It is straightforward to verify that \blue{any valid choice of covariance function $C$ on $\mathbb R^d$ will yield a positive definite covariance function $C^*$ on $\calD$ and that $w(\cdot) \sim GP(0, C^*)$ on $\calD$ is well-defined %
(in the sense of Kolmogorov's conditions). We give the explicit expression of $C^*$ below. %
For notational convenience, for $s \in \calV$, we define $N(s) = \{s\}$, $B(s) = 1$, and $F(s) = 0$. Then %
(\ref{eq:process}) holds for all $s \in \calD$. We then have 
\begin{equation}\label{eq:covfunc}
C^*(s,s') = Cov(w(s), w(s')) = B(s)L(N(s),N(s'))B(s') + \delta(s,s') F(s)\,\,  \forall s,s' \in \calD,
\end{equation}
where $L=CovSel(C(\calV,\calV),\calG)$ as defined in (\ref{eq:finite}) and $\delta(s,s')=I(s=s')$.} The construction \blue{of $C^*$} only relies on the parent \blue{Euclidean} covariance function $C$ and the visibility graph $\calG$. %
The parameters of $C^*$ are thus the same as those of $C$. 

We refer to a process \blue{$w(\cdot) \sim GP(0,C^*)$} as \textit{visGP} due to its reliance on the visibility graph. 
\blue{Specific constructions of neighbor sets are discussed in Section \ref{sec:prediction}. The ``nearest clique" strategy adds one neighbor at a time until the sub-graph would no longer be complete, the ``maximum precision" strategy finds a neighbor set whose implied precision for the new prediction is highest. %
With small sample size or in areas of the domain with low connectivity, one might select $k$ to be the total number of observations connected to $s$ and each other, creating a maximal neighbor set. Otherwise, $k$ may be selected to be some fixed number for computational convenience. Previous research on NNGP has found that using values as low as 10-15 can yield good performance \citep{datta2016}.}

\subsection{Properties}\label{sec:theory}
As discussed in Section \ref{sec:finite}, our visibility graph-based approach is motivated by two principles. The first is that the analysis restricted to any convex subset of the domain should correspond to a traditional geospatial analysis on a convex domain. This translates to preserving all the marginal distributions and pairwise covariances among points connected in the domain. The second is that the conditional covariance of points not connected in the domain is zero. This is intuitive, as the domain boundaries can be viewed as preventing any direct information flow between the two points that can result in conditional correlation. The construction of the process (\ref{eq:finite}) on the finite set $\calV$ using covariance selection immediately guarantees these properties hold on $\calV$, \blue{as discussed after Equation (\ref{eq:finite}).} The following \blue{two} results show that the extension to a process $w(\cdot)$ on the entire domain $\calD$, achieved via (\ref{eq:process}), retains these properties. \blue{We first state and prove an exact result that the process construction preserves marginal distributions %
as specified by the parent covariance function $C$. 

\begin{proposition}\label{prop:marg} 
Consider any finite set of locations $\calV$ in a partially Euclidean non-convex domain and let $\calG$ denote the visibility graph on $\calV$ based on $\calD$. Let $C$ denote any valid Euclidean (isotropic) covariance function on $\mathbb R^d$ and $C^*$ denote the visGP covariance function (\ref{eq:covfunc}) on $\calD$ derived from $C$ and $\calG$. 
Then %
the visGP $ w(\cdot) \sim GP(0,C^*)$ satisfies:\\
$\mbox{(Marginal stationarity:)}\qquad  w(s) \overset{d}{=} w(s') \mbox{ for any } s,s' \in \calD.$
\end{proposition}
}

Proofs of all the theoretical results are provided in Section \ref{web_app_B}. \blue{Proposition \ref{prop:marg} shows that the visGP covariance function (\ref{eq:covfunc}) exactly preserves the marginal variances of the parent covariance function $C$. The result is exact and holds for any finite $\calV \in \calD$ and any valid isotropic $C$. %
This proves the first of the three properties (marginal stationarity) for visGP at the process level (i.e., on the entire $\calD$). The two other properties (preservation of covariances for points connected in the domain, and Markov on points not connected in the domain) hold exactly on $\calV$ but asymptotically on $\calD \setminus \calV$ as proved in the following result.}

\begin{proposition}\label{prop:process} Consider an increasing collection of finite locations $\calV_n$ in a \blue{partially Euclidean} non-convex domain $\calD \in \mathbb R^d$ such that $\cup_n \calV_n$ is dense in D. Let $\calG_n$ be the visibility graph on $\calV_n$ based on $\calD$. Let $C$ be any valid Euclidean (isotropic) covariance function on $\mathbb R^d$ and $C^*_n$ denote the covariance function of the visGP 
$w(\cdot) \blue{\sim GP(0,C^*_n)}$ on $\calD$ %
using $\calV_n$ and $\calG_n$ and with neighbor sets $N(s)$ described in Section \ref{sec:process} \blue{satisfying} $\|B(s)\| \leq M$ for some $M$ for all $s \in \calD$. \blue{Also, for any two locations $s,s' \in \calD$, define the conditional visGP covariance %
$C^*_n(s,s' \given \cdot) := Cov\Big(w(s),w(s') \given \big\{w(u) \given u \in \calD \setminus \{s,s'\}\big\} \Big).$}
Then we have the following:\\
 (Partially Euclidean:) $\lim_n \blue{C^*_n(s,s')} = C(s,s')$ for any $s,s'$ %
    connected in $\calD$, and \\
(Markov:) $\blue{C^*_n(s,s' \given \cdot)}  =0$ for large enough $n$ for any $s,s' \in \calD$ that are not connected in $\calD$.
\end{proposition}
We note that Proposition \ref{prop:process} requires minimal assumptions. It enforces no restriction on the shape of the domain or on the choice of the parent covariance function $C$ beyond isotropy, or on the design of the finite set of locations $\calV_n$ (which in practice is typically the set of data locations). Thus, irregular data designs are accommodated, with the asymptotic regime assuming that data locations will become dense in $\calD$. %
The condition $\|B(s)\| \leq M$ bounds the kriging weights $B(s) = C(s,N(s))C(N(s),N(s))^{-1}$. In less technical terms, this essentially prohibits the neighbors from being chosen very close to each other, as then the contributions by the different members of $w(N(s))$ in predicting $w(s) \given w(N(s))$ become less identifiable and the kriging weights can diverge. We note that this assumption is purely on the construction of the neighbor sets, which is controlled by the user and can be enforced by sequentially choosing neighbors that are sufficiently distant from the previously chosen neighbors.

\blue{Propositions \ref{prop:marg} and \ref{prop:process}} prove that visGP %
covariance function $C^*$ %
satisfies desirable properties at the process level on the entire non-convex domain $\calD$. \blue{Proposition \ref{prop:marg} and the partially Euclidean property of Proposition \ref{prop:process}} ensure that the covariance function $C^*$ restricted to any convex subset $\calD_c \subset \calD$ agrees  with $C$, thereby preserving marginal stationarity \blue{exactly} on all of $\calD_c$ and Euclidean distances \blue{exactly on $\calV \subset \calD_c$ and asymptotically on the rest of $\calD_c$}. It thus ensures that any sub-analysis of the data using $C^*$, restricted to a convex subset $\calD_c$, is equivalent to an analysis using the Euclidean distance-based covariance function $C$. The Markov property \blue{of Proposition \ref{prop:process}} ensures conditional independence between two points not connected in $\calD$. This encodes the irregular non-convex geometry of the domain, as the correlation between the process realizations at these two points is likely to come from correlations of each of them with realizations at intermediate locations in the domain. %
\blue{The following result illustrates, for a class of non-convex domains, how the visGP covariance function is exactly informed by the %
non-convex geometry via the Markov property
.}

\begin{proposition}\label{prop:convex_union} Let $\calD \subset \mathbb R^d$ denote an irregular simply connected domain equipped with the geodesic distance $d_{geo}$ such that $\calD = \cup_{i=1}^J A_j$ where $A_j$'s are convex, and $A_j \cap A_{j'}$ is either empty or contains a single location $s_{jj'}$. Let $C$ denote the exponential covariance on Euclidean distance in $\mathbb R^d$, i.e., $C(s_i,s_j) = \sigma^2 \exp(- \phi \|s_i - s_j\|)$ for $s_i, s_j \in \mathbb R^d$. Let $C^*$ denote a visGP constructed using a finite set of locations $\blue{\calV} \subset \calD$ that contains all $s_{jj'}$. Then $C^*$ \blue{is exactly $C$ with geodesic distances, i.e., } $C^*(s,s')=\sigma^2\exp(-\phi\; d_{geo}(s,s'))$ for all $s,s' \in \blue{\calV}$. 
\end{proposition}

\blue{Proposition \ref{prop:convex_union} is an exact result and} proves that for domains that can be represented as the union of convex domains touching at at-most a single point, a visGP constructed from a parent GP with an exponential covariance function with Euclidean distance has an exponential covariance function with the geodesic distance on the non-convex domain. \textcolor{black}{Figure \ref{web_fig_3} of the supporting information} provides examples of domains that can be characterized in this way, including tree-shaped domains (left) and unions of polygons (right). %
For these domains, the geodesic distances are exactly encoded in the visGP exponential covariance, demonstrating how the Markov property on the visibility graph incorporates the domain geometry. 
We show in Section \ref{sec:simcov} that this property holds approximately even in domains excluded from the premise of \blue{Proposition \ref{prop:convex_union}.}
Note that %
an immediate corollary is \blue{that the exponential covariance function is positive-definite on the} non-convex domains considered in \blue{Proposition \ref{prop:convex_union}}, which is a result of independent importance. %

\section{Computational strategies}\label{sec:approx}
We now outline the algorithm to analyze geospatial data on non-convex domains using visGP and provide strategies for improving scalability. Consider \blue{a univariate response} $\blue{Y_i :=} Y(s_i)$ and \blue{a $p$-dimensional covariate} $X_i \,\blue{:= X(s_i)}$ observed at locations $s_i$, for $i=1,\ldots,n$ in a non-convex \blue{partially Euclidean} domain $\calD \subset \mathbb R^d$. %
We consider $\calV$ to be the set of data locations and define the visibility graph $\calG$ on $\calV$ as in Section \ref{sec:finite}. Note that if the data locations leave large gaps in the domain, one can always add more points to $\calV$ and define $\calG$ on this augmented set of locations. Let $\Sigma(\cdot,\cdot)$ a parent Euclidean covariance function on $\mathbb R^d$ that combines a spatial GP with Euclidean covariance $C(\cdot,\cdot)$ and a noise (nugget) process $\epsilon(s) \iid N(0,\tau^2)$. 
Let $w(\cdot)$ denote a visGP with covariance function $\Sigma^*$ based on $\Sigma=C + \tau^2 \delta$ where $\delta(s,s')=I(s=s')$. %
Then the visGP process model is given by $Y(s) = X(s)'\beta + w(s)$, \blue{$w(\cdot) \sim GP(0,\Sigma^*)$}.
Defining $Y=(Y_1,\ldots,Y_n)'$ and $X$ similarly, we have 
\begin{equation}\label{eq:model}
    Y = N\Big(X\beta,\Sigma^*(\calV,\calV) \Big) \mbox{ where } \Sigma^*(\calV,\calV)=CovSel\big(C(\calV,\calV) + \tau^2 I,\calG\big).
\end{equation}

The parameters of the visGP covariance $\Sigma^*$ are simply the parameters $\theta$ of the original GP $C=C(\theta)$ and the nugget variance $\blue{\tau^2}$. Given these, the matrix $\Sigma^*(\calV,\calV)$ can be calculated using the iterative proportional scaling (IPS) procedure of \cite{speed}. Hence, all parameters $(\beta,\theta,\tau^2)$ can be estimated by maximizing the likelihood corresponding to (\ref{eq:model}).

For moderate to large sample sizes,  the IPS algorithm can be computationally intensive as it involves an iterative procedure. %
We propose a few approximations which preserve the spirit of the method while minimizing computational overhead. %

\subsection{Chordal completion}
We first consider computations for the setting where the visibility graph $\calG$ is a \textit{chordal} or \textit{decomposable} graph. A graph $\calG$ is said to be chordal if every one of its cycles of length four or greater has a chord. In graphical statistics, chordal graphs have attractive computational properties. We make use of the following from \cite{lauritzen}.
The maximal cliques (i.e., complete sub-graphs which are not contained in larger complete sub-graphs) of a chordal graph $\calG$ admit a perfect ordering $(K_1, K_2, ..., K_{\blue{c}})$, i.e., one where we can write
\[
H_{j} = K_1 \cup ... \cup K_{j},\,
R_j = K_j \setminus H_{j-1},\,
\blue{S}_j = H_{j-1} \cap K_j
\]

and $(H_{j-1}, R_j, S_j)$ is a \textit{decomposition}, meaning $S_j$ \textit{separates} $H_{j-1}$ from $R_j$; i.e., all paths from any vertex in $H_{j-1}$ to any vertex in $R_j$ goes through $S_j$. Hence, $S_j$ are referred to as ``separators," and since they are sub-graphs of cliques, they are themselves cliques. For such a perfect ordering for the visibility graph $\calG$, the likelihood for the data model (\ref{eq:model}) is given by 
\begin{equation}\label{eq:chordal}
f(Y \given X, \beta, \theta, \tau^2) = \frac{\prod_ i %
N(Y(K_i) \given X(K_i)\beta, C(K_i,K_i) + \tau^2 I)
}{\prod_i %
N(Y(S_i) \given X(S_i)\beta, C(S_i,S_i) + \tau^2 I)},
\end{equation}
where %
for a set $A \subset \calV$, $Y(\blue{A})$ denotes the subset of $Y$ corresponding to locations in $A$; $X(A)$ is defined similarly, \blue{and $S_1$ is defined as the empty set.} %

The closed-form representation (\ref{eq:chordal}) of the likelihood completely circumvents the IPS algorithm to calculate the covariance matrix $\Sigma(\calV,\calV)$. In fact, the large $n \times n$ matrix $\Sigma(\calV,\calV)$ or its visGP analog $\Sigma^*$ need not be calculated directly at all, as the likelihood decomposes along the smaller clique and separator likelihoods. As the cliques and separators are complete sub-graphs, the corresponding likelihoods for these subsets are simply based on the original Euclidean GP with covariance $C + \tau^2 \delta$. This allows the likelihood to be calculated significantly quicker, as long as the cliques and separators are small relative to the entire graph. 

For certain non-convex domains, the visibility graph is naturally chordal. Examples include  tree-shaped domains (Figure \ref{web_fig_3}, left) %
and rectangular ``U"-shaped domains, like the symbol $\bigsqcup$. 
For some others, like for domains admitting a decomposition of convex domains as in \blue{Proposition \ref{prop:convex_union}}, the graph can be pruned to be chordal by removing edges between points lying in different convex components, while exactly preserving the visGP covariance function. For other domains, we use a \textit{chordal completion}, $\bar\calG$, which is a chordal graph of which $\calG$ is a sub-graph. Intuitively, $\bar\calG$ \blue{adds some edges to $\calG$ to serve as necessary chords.} %
We use a linear-time chordal completion algorithm provided by the \texttt{igraph} software package \citep{csardi}. %
\blue{Section \ref{web_app_D} shows empirically that the chordal completion introduces minimal distortion to the geometry of the domain.}
We replace $\calG$ with the approximate chordal graph $\bar\calG$ for parameter estimation by maximizing the likelihood (\ref{eq:chordal}) based on $\bar\calG$.

\subsection{Graph stochastic gradient descent}\label{sec:sgd}

Note that due to the likelihood decomposition (\blue{\ref{eq:chordal}}) over the cliques and separators, we may write the log-likelihood of our model using a chordal graph as above as
\begin{equation}\label{eq:sgd}
\log f(Y) = \sum_i %
\log f(Y(K_i))
-\sum_i %
\log f(Y(S_i)) = \sum_i[ \log f(Y(K_i)) - \log f(Y(S_i))]
\end{equation}
where $f(Y(A))$ is the likelihood for $Y(A)$. Here we use the fact that each clique $K_i$ in the perfect ordering has a corresponding separator $S_i \subset K_i$. %
Thus, the loss function optimized to obtain parameter estimates is additive over the clique-separator pairs $(K_i,S_i)$ in the perfect ordering of the graph, %
and %
is amenable to maximization by stochastic gradient descent \citep[SGD,][]{shaley}. SGD is a kind of gradient-based optimization in which, at each iteration, the total gradient for an additive loss is approximated by a single component of the loss, and the components are cycled over the iterations. %
In most applications, the ``components" are i.i.d. data points (or blocks). In a spatial setting, all data are correlated, \blue{and the loss functions (log-likelihoods) from spatial models are not additive over data units,} which rules out a naive application of SGD. Instead, here we formulate a novel application of SGD on decomposable graphs, exploiting the additive decomposition (\ref{eq:sgd}) of the log-likelihood where the component is a clique-separator pair. This enables the evaluation of only a couple of Gaussian likelihoods (corresponding to some $K_i$ and $S_i$) at each iteration of the estimation, thereby massively reducing the computation burden. 

Further details of the computational techniques used to expedite the algorithm are provided in Section \ref{web_app_C}. The {\em graph SGD} algorithm is formally given in Algorithm \ref{alg:sgd} of Section \ref{sec:sup_sgd}. In Section \ref{sec:nngp}, we outline nearest neighbor Gaussian process approximations for large clique or separator component likelihoods. In Section \ref{sec:dist}, we discuss how the computational burden can be further eased by introducing, either at the stage of calculating the adjacency matrix or at the stage of likelihood optimization, a distance threshold beyond which two points are considered to be non-adjacent even if they are connected in the domain. %
For predictions  at a new location $s$ using visGP, we can use the kriging equations (\ref{eq:process}) using a neighbor set based on the  %
``nearest clique" %
or ``maximum precision" strategy. %
\blue{We also consider a} ``precision-weighted" \blue{prediction strategy that uses a precision-weighted average over multiple \blue{predictions based on different choices} of neighbor sets  (see Section \ref{sec:prediction}).}

\section{Simulation study}\label{sec:sim}
\subsection{Predictive performance}
\begin{figure}
    \centering
    \begin{subfigure}[t]{.49\textwidth}
        \centering
        \vspace{0pt}
        \includegraphics[scale=.35]{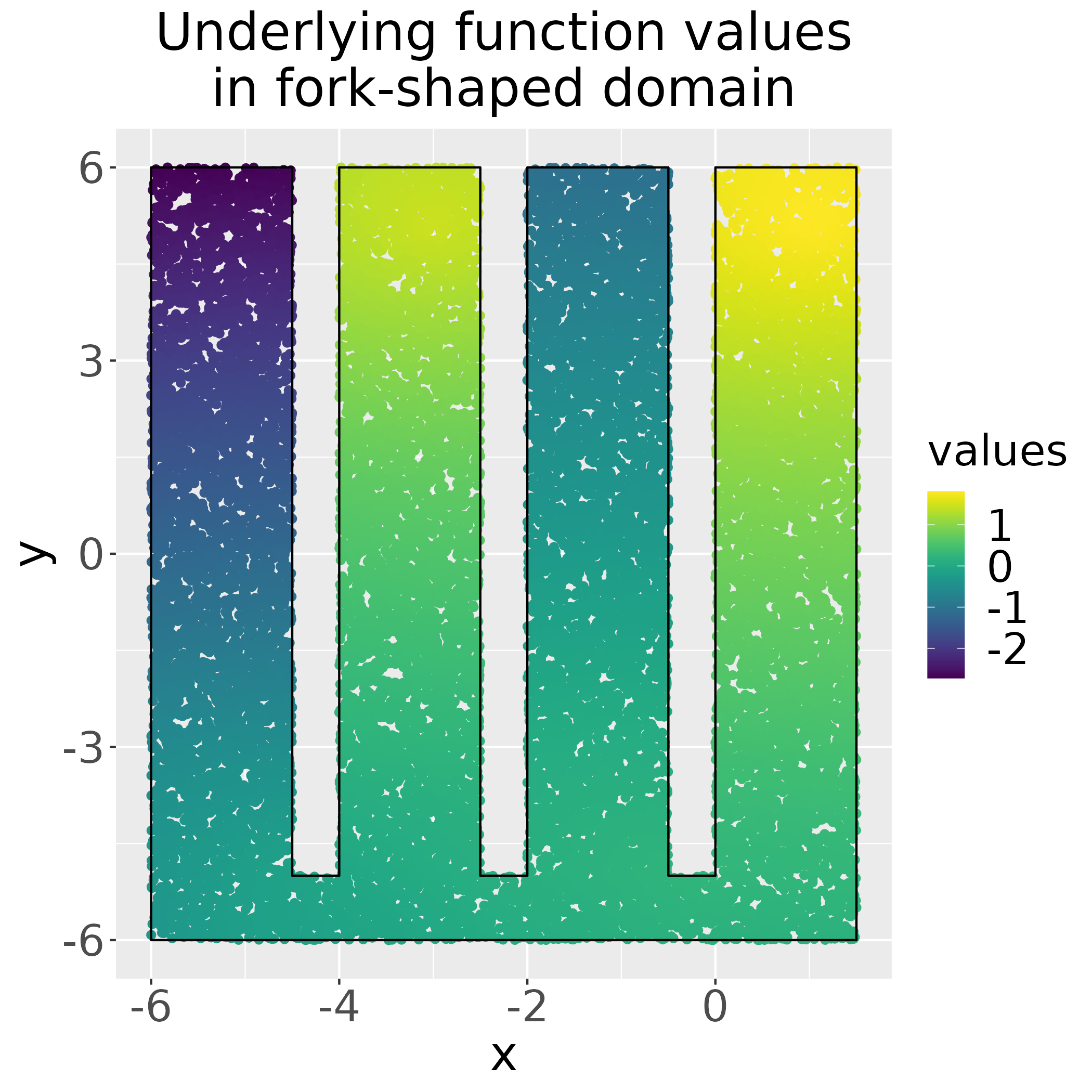}
        \caption{Function values in the fork-shaped domain}
        \label{fig:fork_vals}
    \end{subfigure}
    \hfill
    \begin{subfigure}[t]{.49\textwidth}
        \centering
        \vspace{0pt}
        \includegraphics[scale=.35]{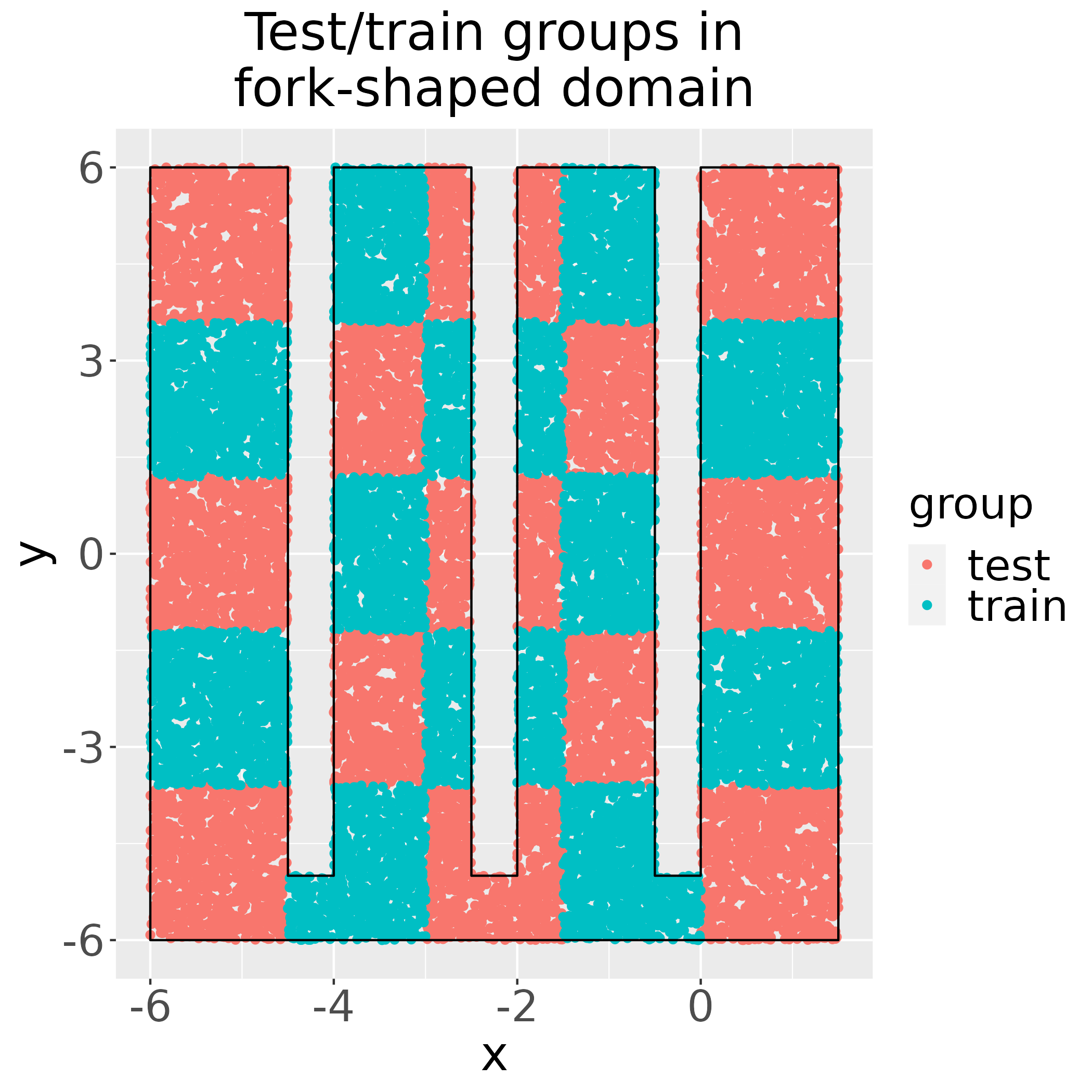}
        \caption{Data grouping in the fork-shaped domain}
        \label{fig:fork_groups}
    \end{subfigure}
    \begin{subfigure}{\textwidth}
        \centering
        \includegraphics[scale=.35]{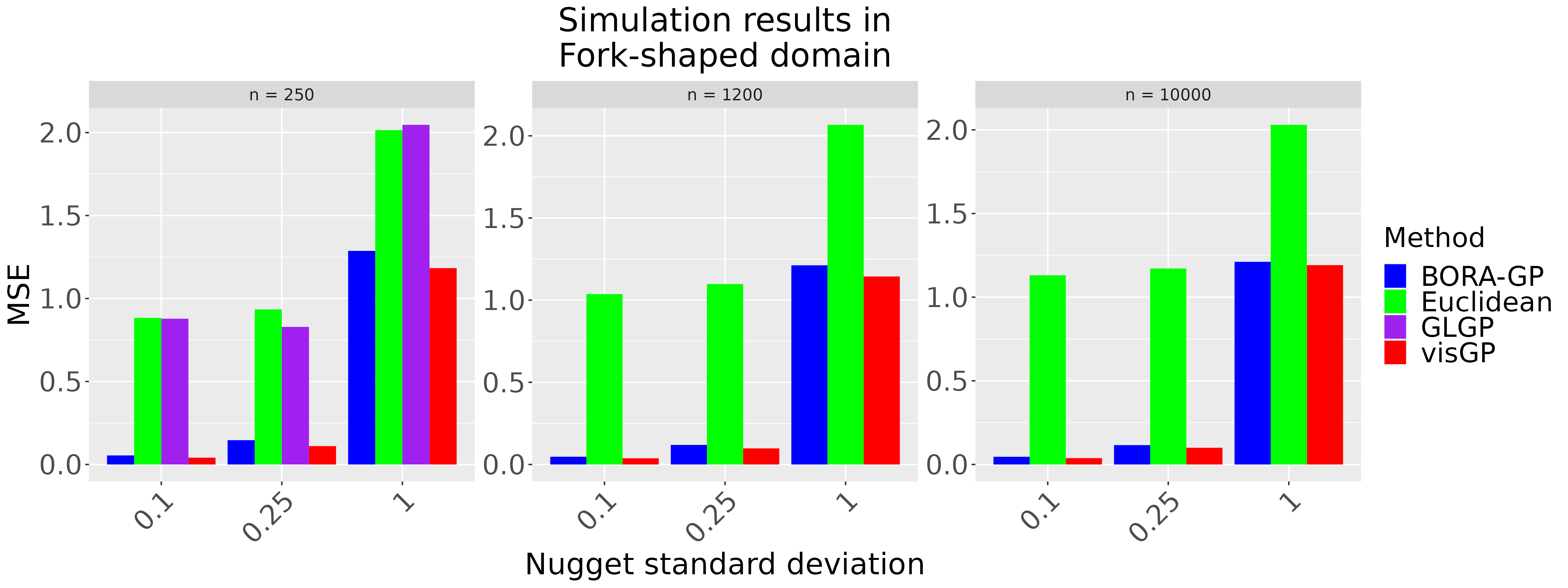}
        \caption{Simulation results in the fork-shaped domain}
        \label{fig:sim_all_res}
    \end{subfigure}
    \caption{Simulation design and results in the fork-shaped domain}
\end{figure}

We examine the performance of various methods by evaluating their predictive accuracy in a synthetic non-convex domain. We use a fork-shaped domain with four rectangular prongs which are spaced parallel to each other and connected by a base region (Figure \ref{fig:fork_vals}). We ensure that for all the simulation settings, our model is misspecified with respect to the true data generation process, i.e., we do not assume the data is generated from a visGP but that there is an underlying fixed spatially-smooth function $f$ that generates the expected value of the spatial process $Y$ at each point and there is a white-noise error variance beyond this function, which is varied across replicate simulation runs, \blue{i.e., $Y(s) = f(s) + \epsilon(s), s\in \calD$ where $f$ is a fixed function and $\epsilon(s)$ are i.i.d. error process in $\calD$.} To create the fixed function \blue{$f$}, we use various ``source" points and calculate through-domain distances, as described \blue{formally in Section \ref{web_app_E}} The values of this function can be seen in Figure \ref{fig:fork_vals}.
We then divide the region into test and training data, as shown in Figure \ref{fig:fork_groups}. For sample sizes of $n=250$, $n=1200$, and $n=10,000$, we divided data into $80 \%$ training and $20\%$ test. To create the values of the spatial process, white noise with standard deviations of $sd=0.1, 0.25, 1$ was added to the underlying functions $f$, described above. \blue{We fit four candidate models --- a GP using Euclidean distances which ignores the water geometry, visGP, BORA-GP \citep{jin2022spatial} and GLGP \citep{dunson}. For details of the specific models fit, see Section \ref{web_app_C}.} Models were fit to the training data, and the point estimates and confidence/credible intervals were compared to the holdout set. Code for the simulations is available at \texttt{github.com/bjg345/visGP}.

\blue{Results are displayed in Figure \ref{fig:sim_all_res} for the visGP and the three competing methods. 
 BORA-GP and visGP tend to provide superior predictions; the Euclidean GP is associated with significantly higher predictive error. This is expected. The GLGP also yields very high prediction errors, which is likely due to the fact that it does not directly incorporate information about the adjacency relations between points, as well as the difficulty of its parameter optimization by grid search.
BORA-GP has somewhat higher MSE than visGP. Tables with full results evaluating both the point estimates and uncertainty estimates for all three versions of visGP, as well as the competing methods, can be found in Section \ref{web_tab_1}. GLGP does not offer prediction intervals and could not be implemented for the larger sample sizes due to computational issues. The predictive intervals for visGP are shorter than those for BORA-GP; both methods attain approximately 95\% coverage, though BORA-GP is occasionally slightly over-conservative and visGP is occasionally slightly anti-conservative. The visGP method presented in Figure \ref{fig:sim_all_res} used the maximum-precision prediction strategy. There are small differences in accuracy between the three prediction strategies for visGP, but none are clearly superior or inferior. %
We note that the differences between methods may be idiosyncratic with respect to the domain under consideration as the likelihood function (with variable mean, spatial range, spatial variance, and nugget variance) is overparametrized for a domain of fixed diameter \citep[see][]{zhang}. Also, many predictions rely on extrapolation due to the checkerboard pattern of the holdout set, as seen in Figure \ref{fig:fork_groups}. 

To assess the performance of the methods in a setting that does not require much extrapolation, a supplementary experiment performed on a random, dense holdout set, presented in Section \ref{web_tab_2}, shows both visGP and BORA-GP attain nominal coverage, with visGP tending to have lower MSE, especially for smaller sample sizes and lower nugget variance. In addition, a further experiment in a U-shaped domain with a dense holdout set, described in Section \ref{web_app_F}, shows the tendency of visGP to estimate a range parameter value that is higher relative to its variance than for BORA-GP. This would explain its shorter intervals, even though both visGP's and BORA-GP's estimated parameters better optimize the likelihood under their respective models. }

Finally, for a randomly selected simulation run \blue{in the U-shaped domain} with $n=10,000$, we compared the \blue{total} runtimes of the BORA-GP and visGP, starting from the raw data on the domain shape, locations, and observed values to final test-set predictions. We used R version 4.0.3 and a local machine [Intel(R) Core(TM) i5-8265U CPU @ 1.60GHz-1.80 GHz] running Windows 10 x64. The results can be seen in Table \ref{tab:timetest}.

\begin{table}[ht]
\caption{Computation times for the visGP and BORA-GP methods. \blue{visGP-fast denotes a faster implementation of visGP where the adjacency calculation for the visibility graph is restricted to be within a distance threshold right at the onset.} Units are minutes.}\label{tab:timetest}
\centering
\begin{tabular}{lcccc}
\hline
Method & Neighbor-finding & Model-fitting & Prediction & Total\\
\hline
BORA-GP & 11.43 & 90.76 & 10.00 & 112.19\\
visGP   & 68.79 & 16.64 & 0.08 & 85.51 \\
\blue{visGP-fast}  & 3.48 & 16.64 & 0.08 & 20.20
\\ \hline
\end{tabular}
\end{table}

\blue{In total, the BORA-GP method took $112.19$ minutes, while the visGP method took %
$20.20$ minutes if the adjacency matrix was thresholded at its creation.} The large majority of the computation time for BORA-GP was in model-fitting, while the large majority of the computation time for visGP \blue{without thresholding} was in constructing the adjacency matrix, which is a ``one-time" cost. Multiple analyses (e.g., at different time-points) on the same domain can be accomplished without having to recalculate the adjacency matrix. %
The problem is also embarrassingly parallel, as every pair's adjacency can be calculated independently. The time required for fitting the model for visGP was more than $5$ times faster than that of BORA-GP. 

\subsection{Process properties}\label{sec:simcov}

\begin{figure}[!t]
    \centering
    \begin{subfigure}[t]{.49\textwidth}
    \includegraphics[scale=.383]{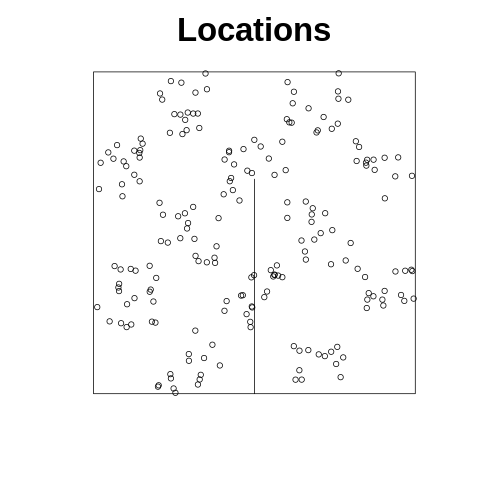}
    \caption{Locations used for the covariance comparison}
    \label{fig:covcomp_locs}
\end{subfigure}
    \begin{subfigure}[t]{.49\textwidth}
         \includegraphics[scale=.383]{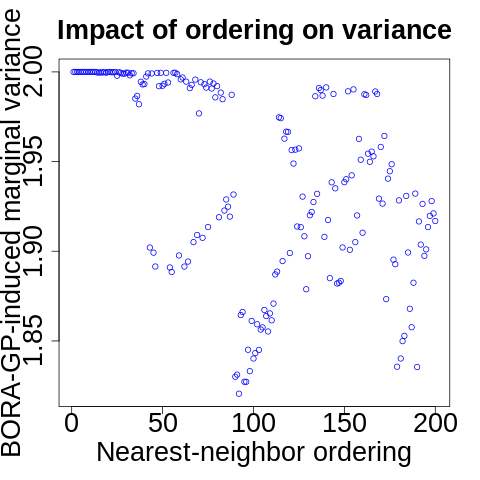}
    \caption{Impact of ordering on the induced marginal variance of BORA-GP.}
    \label{fig:covcomp_ord}
    \end{subfigure}\\
    \begin{subfigure}[t]{.49\textwidth}
    \centering
    \includegraphics[scale=.383]{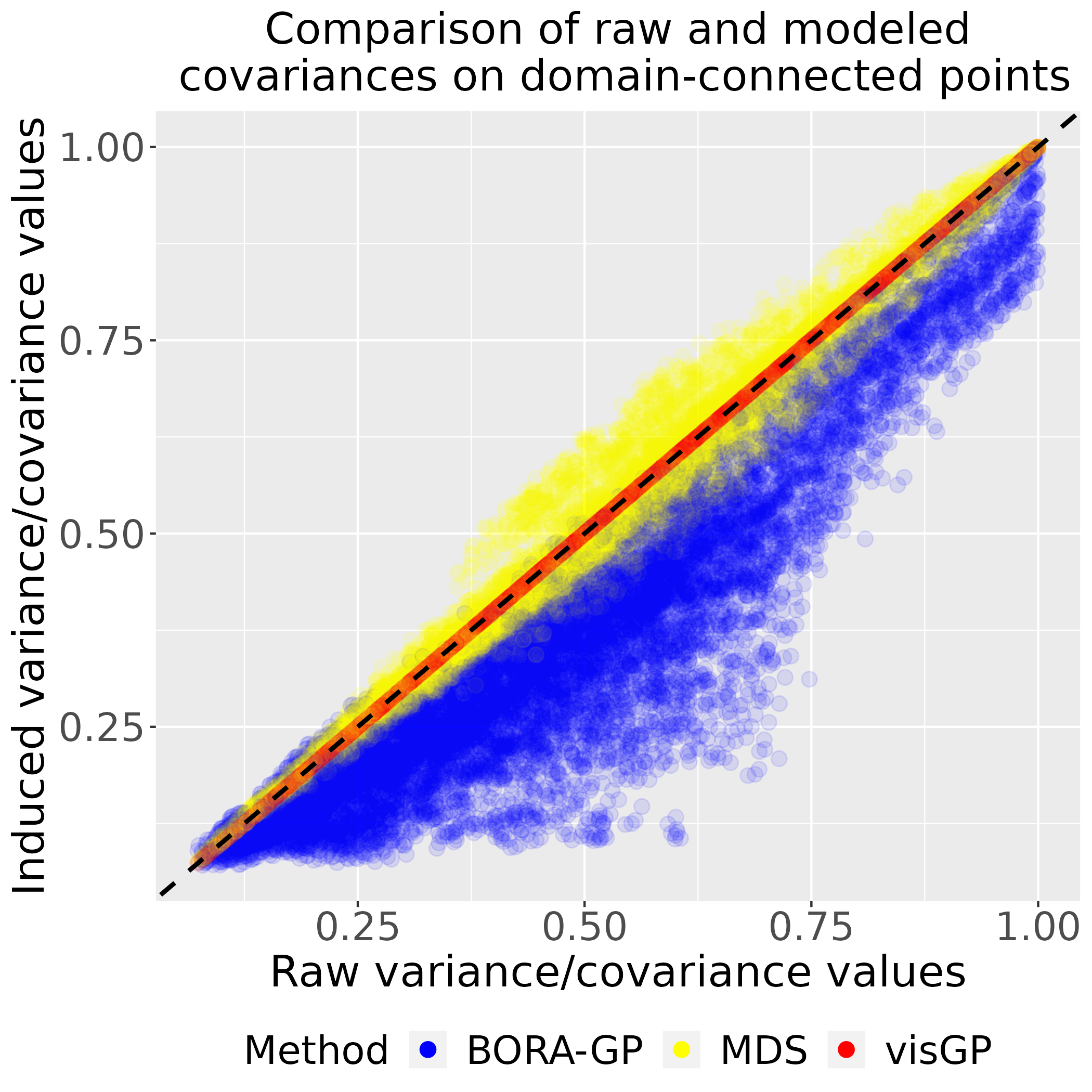}
    \caption{Covariance values comparing the BORA-GP, MDS, and visGP models to the Euclidean values on domain-connected points}
    \label{fig:covcomp_vars_water}
    \end{subfigure}
     \begin{subfigure}[t]{.49\textwidth}
    \centering
    \includegraphics[scale=.383]{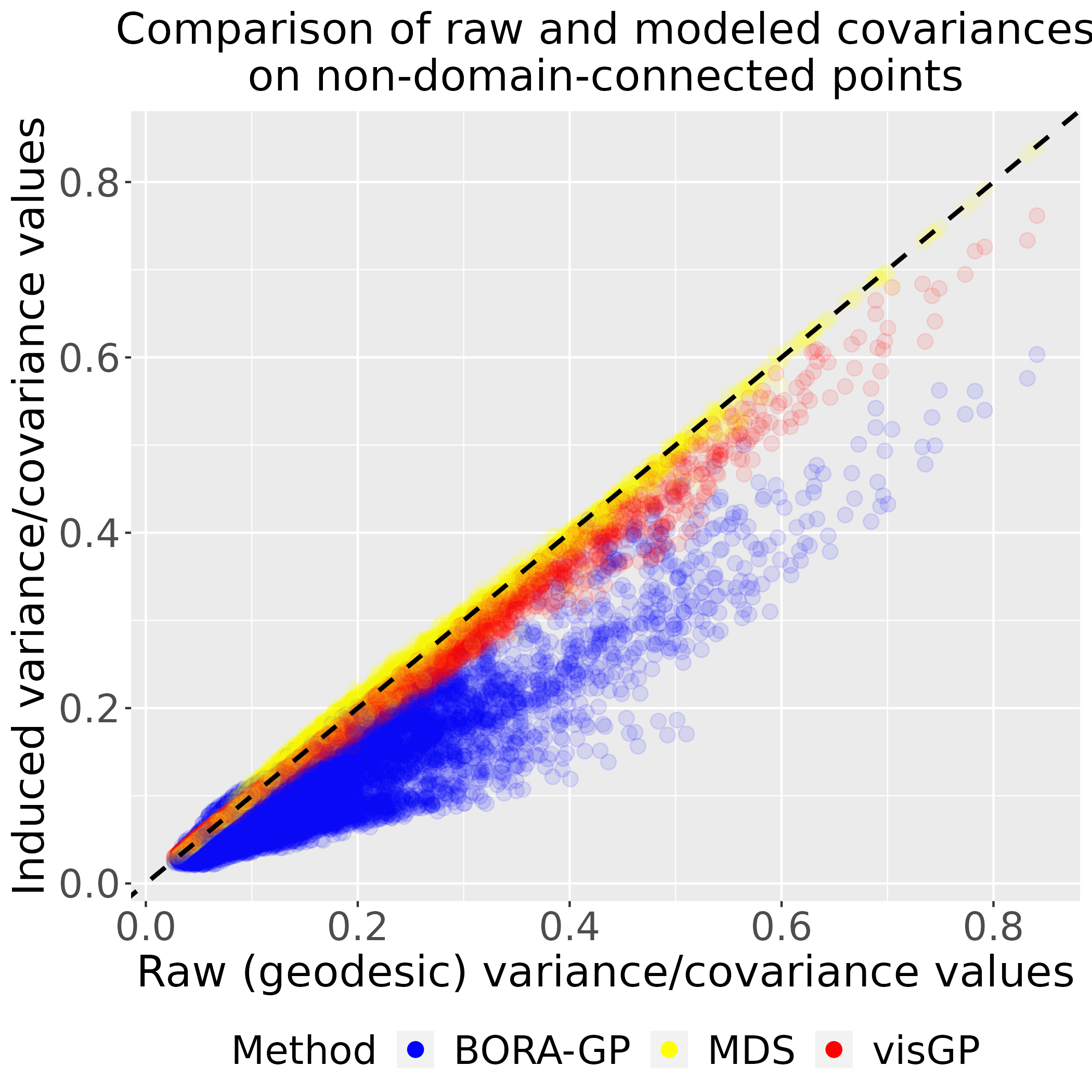}
    \caption{Covariance values comparing the BORA-GP, MDS, and visGP models to the geodesic values on non-domain-connected points.}
    \label{fig:covcomp_vars_land}
    \end{subfigure}
    \caption{Results of the variance-covariance study comparing Euclidean/geodesic, BORA-GP, MDS, and visGP values.}
\end{figure}

As proved in Section \ref{sec:theory},  visGP preserves entries in the covariance matrix which correspond to points connected in the domain. In the following experiment, we demonstrate the extent to which this property is violated by \blue{competing methods.} We consider \blue{locations in} a $U$-shaped domain (Figure \ref{fig:covcomp_locs}) %
and use a parent Euclidean covariance based on the Mat\'{e}rn function with spatial variance $\sigma^2=1$, smoothness $\nu =1$, inverse-range $\phi = 0.1$, \blue{and nugget variance $\tau^2=1$.} We compute the induced variances and covariances for \blue{visGP, BORA-GP, and MDS (multidimensional scaling of the geodesic distance matrix and applying a Euclidean GP). See Section \ref{web_app_H} for details on the data generation and implementation of the methods.} %

We first look at the \blue{total marginal} variances \blue{in Figure \ref{fig:covcomp_ord}. The visGP marginal variances are guaranteed to be $\sigma^2 + \tau^2=2$.} For BORA-GP, the ordering imposed to create the directed nearest neighbors strongly influences the marginal variances, with locations appearing later in the ordering having a decrease in the induced marginal variance. %

 We next compare covariances %
 with the raw values of a Mat\`ern function on geodesic distance. 
 When points are connected in the domain (Figure \ref{fig:covcomp_vars_water}), the geodesic distance corresponds to the Euclidean distance, and the covariances from visGP for these points are exactly identical
to the Mat\'ern covariances on these Euclidean distances. This property of visGP, again, is guaranteed from \blue{the property of covariance selection (see discussion following Equation \ref{eq:finite}),} and together with preservation of the variances, ensures that a visGP analysis restricted to a convex subdomain coincides with standard GP analysis using Euclidean covariances. For BORA-GP \blue{and MDS,} the deviation from the 45-degree line indicates a discrepancy from Euclidean covariances on points connected in the domain, and we see that these deviations are often quite large \blue{for BORA-GP, with a systematic weakening of the covariances.} 

Finally, in Figure \ref{fig:covcomp_vars_land}, we look at covariances for points not connected in the domain. \blue{Proposition \ref{prop:convex_union}} has shown that for certain domains (as in Figure \ref{web_fig_3}) and choice of covariance function (exponential), the visGP covariance is exactly the covariance using the geodesic distance. However, this will not hold exactly in general for \blue{arbitrary partially-Euclidean domains like the $U$-shaped domain and for other covariance functions like the Mat\'ern($\nu=1$) covariance 
 considered here.} %
However, we see from Figure \ref{fig:covcomp_vars_land} that \blue{MDS and visGP methods} retain this property approximately in other domains and other types of covariance functions. The \blue{MDS and visGP} covariances are quite close to the covariance using the geodesic distance. This reflects how the geometry of the domain is embedded into the visGP construction. Once again, for BORA-GP, we see the association with the geodesic covariances is considerably weaker, demonstrating a loss of knowledge about the domain geometry to a greater extent. These figures show that, unlike existing methods, the visGP model exactly preserves Euclidean covariances on domain-connected points and roughly preserves geodesic covariances of non-domain-connected pairs. 
 The former is clear by the design of the visGP method; the latter observation is not obvious from \blue{the Markov} property but is somewhat predicted by the result of \blue{Proposition \ref{prop:convex_union}.} 

\section{Application: acidity of the Chesapeake Bay}\label{sec:chesa}

The Chesapeake Bay is the ``largest, most productive, and most biologically diverse estuary in the United States," according to The Chesapeake Bay Program. Formally founded in 1983, the program aims to protect and restore the Chesapeake Bay and its watershed through ecological monitoring and management in the face of human population growth and environmental degradation \citep{hood2021chesapeake}.
One variable tracked by the Project's monitors is pH, which measures local acidity. pH level has been argued to be an important factor in maintaining an estuary system's biological health \citep{ringwood}. %
In the analysis below, we examine average pH levels measured at each of 213 monitoring locations throughout the year 2021, which can be accessed at \texttt{https://data.chesapeakebay.net/}. Code for our analysis is available at \texttt{github.com/bjg345/visGP}.

\begin{figure}[t]
    \centering
    \begin{subfigure}{0.49\textwidth}
        \centering
        \includegraphics[scale=0.4]{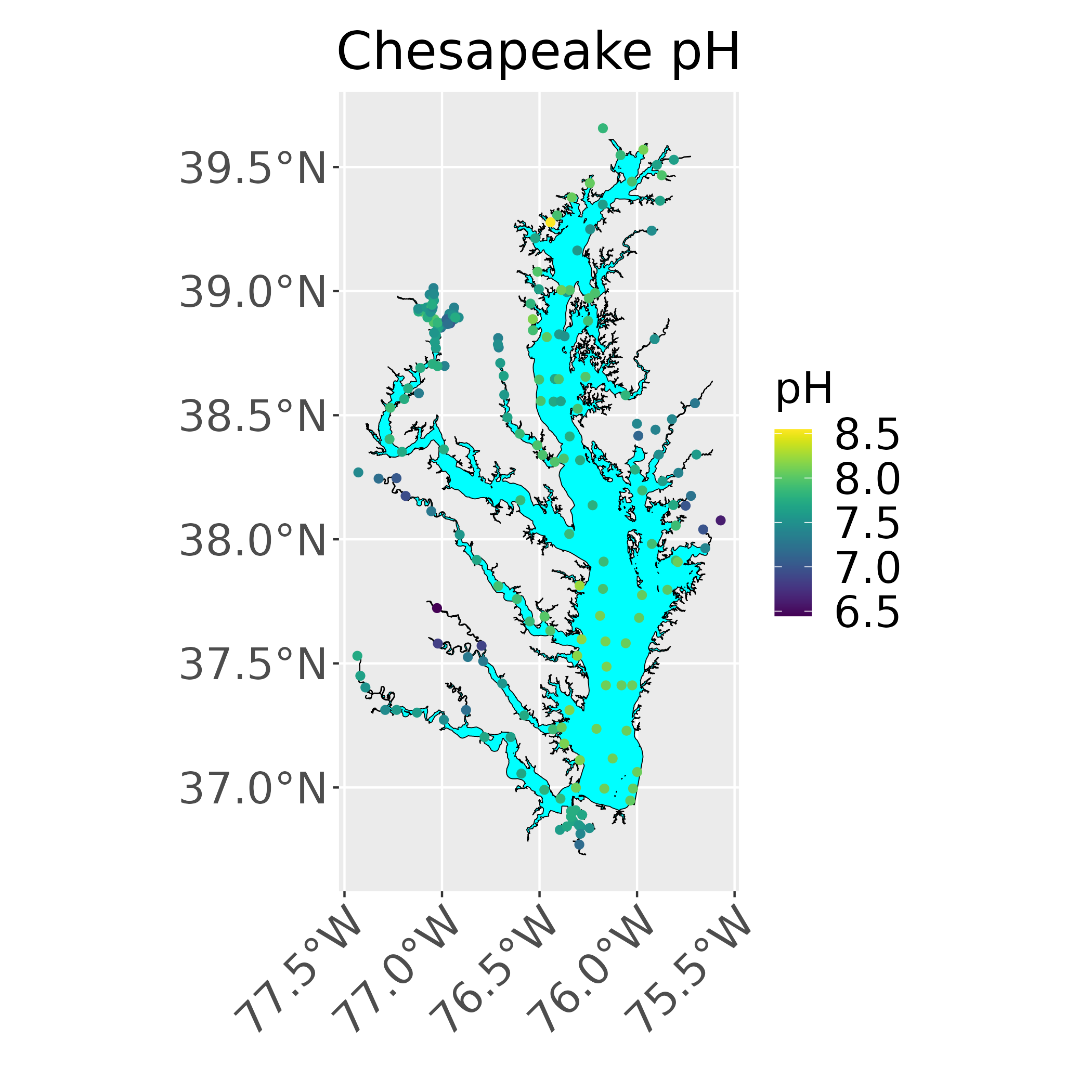}
        \caption{Average pH levels at each monitoring station in the year 2021}
        \label{fig:pH}
    \end{subfigure}%
    \begin{subfigure}{0.49\textwidth}
        \centering
        \includegraphics[scale=0.4]{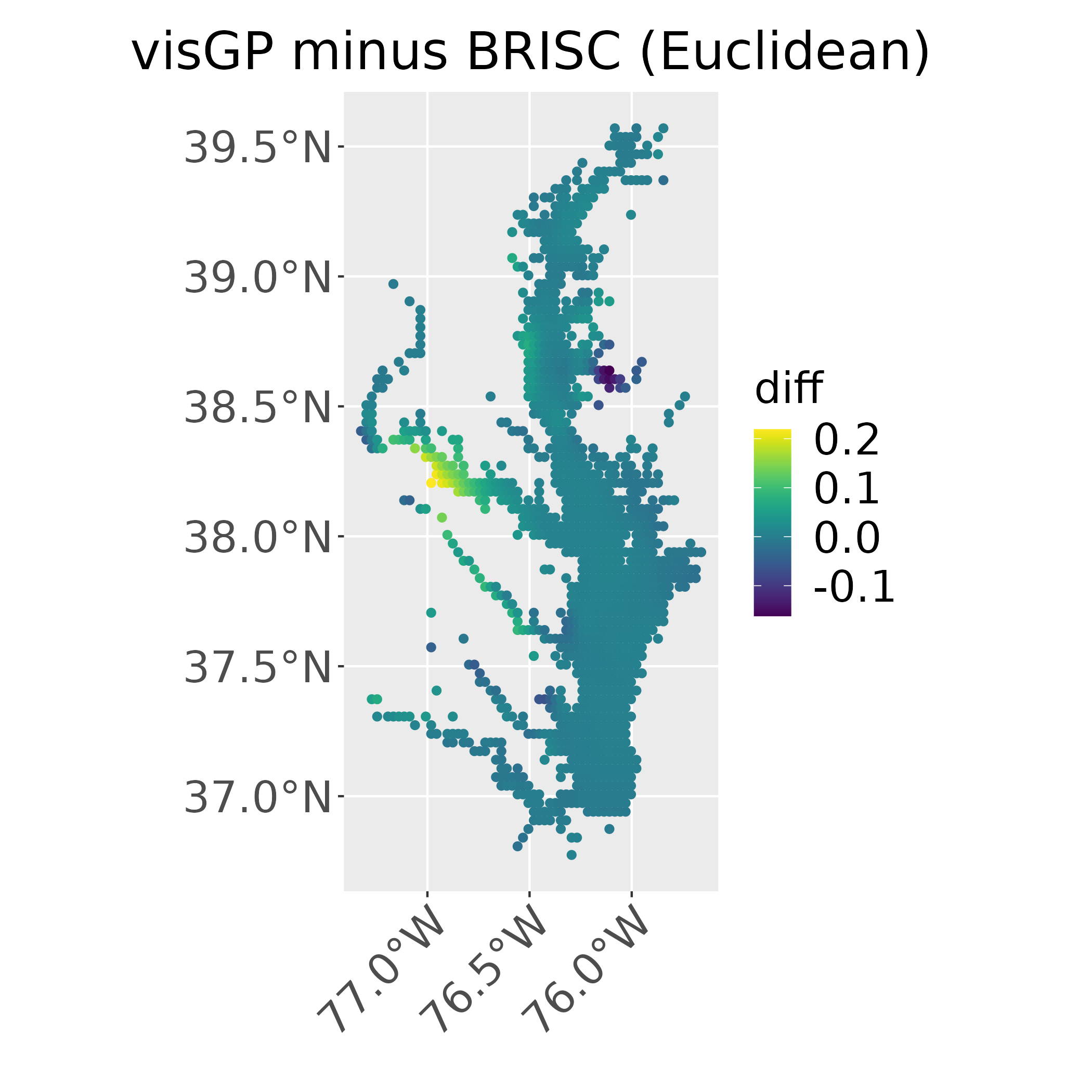}
        \caption{Difference between visGP and Euclidean predictions in the Chesapeake Bay}
        \label{fig:diff}
    \end{subfigure}
    \newline
    \begin{subfigure}{\textwidth}
        \centering
        \includegraphics[scale=0.4]{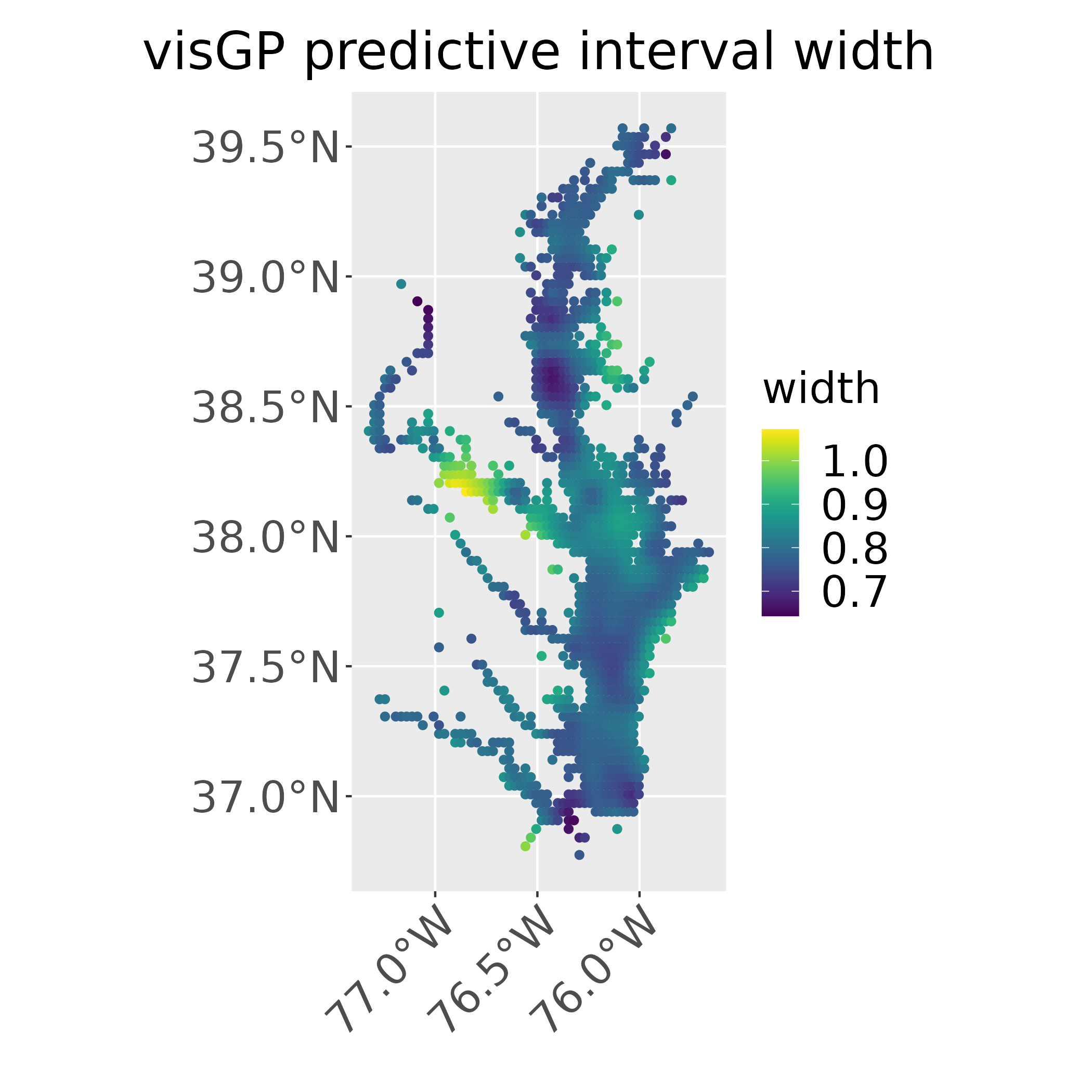}
        \caption{Length of predictive intervals for visGP model}
        \label{fig:predict}
    \end{subfigure}
    \caption{Visualizations of pH levels in the Chesapeake Bay}
\end{figure}
It is apparent that pH levels track the bay's complicated geometry (see Figure \ref{fig:pH}) with considerable variability in levels from different \blue{tidal tributaries} that are close in Euclidean distance but far away in the geodesic or water distance. We compare the performance of three models for predicting the pH levels in this water body -- Euclidean GP (fit by BRISC R-package \cite{saha} with the exponential covariance function and 15 neighbors) which ignores the water geometry, visGP (with the exponential covariance function and maximum-precision clique prediction with 15 neighbors), and BORA-GP model with 15 neighbors. \blue{Section \ref{sec:phdetails} details the implementation of the methods.}

\blue{
To get a sense of model performance, we do fitting and prediction with three-fold holdout sets (using two folds at a time to train the model and predict on the third). These sets are constructed randomly with equal size out of the 213 observations.
The results for the three models are shown in Table \ref{tab:combined} (left). visGP and BORA-GP show improvements over the Euclidean method in terms of MSE. All methods tend to undercover. %

To investigate the under-coverage of the methods further, we do a leave-one-out analysis, which results in larger training sets of size $n=212$. As the leave-one-out analysis needs to be conducted $213$ times, once for each held-out location, we could not implement BORA-GP due to its longer run times. Based on the 3-fold analysis, we expect the BORA-GP results to be very similar to that of visGP. Table \ref{tab:combined} (right) compares visGP and the Euclidean GP for the leave-one-out analysis. We see that, compared to the 3-fold analysis, there is a decrease in MSE for each method, which is expected given the larger training data. However, the gap in prediction accuracy between visGP and Euclidean GP persists even when using larger training sets. Also, both methods now have near-nominal coverage, implying that the under-coverage of the 3-fold analysis was likely due to inadequate training sample size. 

Overall, these results suggest that the assumptions underlying the visGP method fit well with the natural processes governing acidity levels in this domain and can be used to identify or predict areas of concern for protection or intervention, although uncertainty quantification can be problematic for all methods at low sample sizes. }

\begin{table}[ht]
\caption{Combined Results of Chesapeake Bay pH Data Analysis. Here, visGP uses the `maximum precision' strategy to define the neighbor sets. }\label{tab:combined}
\centering
{\color{black}
\begin{tabular}{rllrllr}
  \hline
 & \multicolumn{3}{c}{3-fold Validation} & \multicolumn{3}{c}{Leave-One-Out Analysis} \\
 \cline{2-4} \cline{5-7}
 & MSE & Coverage & CI.length & MSE & Coverage & CI.length \\ 
  \hline
Euclidean & 4.97e-02 & 68.8\% & 0.84 & 3.74e-02 & 94.4\% & 0.81 \\ 
visGP & 4.76e-02 & 66.1\% & 0.81 & 3.46e-02 & 94.4\% & 0.78 \\ 
BORA-GP & 4.75e-02 & 67.9\% & 0.84 & - & - & - \\ 
  \hline
\end{tabular}
}
\end{table}

\blue{
To understand the importance for a statistical analysis to be informed by the geometry of the bay, we consider the differences between visGP and Euclidean GP predictions, conditional on training on all observations. For creating maps, the predictions are interpolated on a fine grid %
and shown in Figure \ref{fig:diff}. The most notable difference is in the region around \( 38.25^\circ \text{N}, 77.00^\circ \text{W} \) in the Potomac River. Here, the Euclidean GP predicts lower pH values because it uses neighbors from the Rappahannock River (the tributary below), which is close in Euclidean distance. However, the tributaries are farther apart in geodesic distance and are not connected in the visibility graph for the data locations.  Hence, data from the Rappahannock River are not used for visGP predictions in that area of the Potomac River. The results from visGP align with the actual pH levels measured in the stations where the pH levels in the Potomac seem to be distinctly higher than those in the Rappahannock River. The results are also supported by published scientific literature on differences in water chemistry in different tidal tributaries of the Chesapeake Bay. In general, variables like alkalinity and salinity, which influence the pH levels, have been shown to vary widely between the Potomac and Rappahannock rivers and are dictated by the characteristics of the non-tidal rivers feeding into these tributaries \citep{najjar2020alkalinity}. Relative to the Rappahannock River, higher pH levels in the Potomac River, have also been observed in \cite{da2021mechanisms}. Finally, Figure \ref{fig:predict} plots the predictive uncertainty of pH levels across the bay, from visGP. It indicates that this segment of the Potomac River is also the area with the most predictive uncertainty, suggesting a need for additional monitoring in that area.}

\section{Conclusion}\label{sec:}
Inference and prediction for spatial processes in non-convex \blue{partially Euclidean} domains are often encountered in practice. %
Many methods \blue{for such analyses} have taken the differential equation perspective to construct Gaussian processes in these domains, \blue{and none respect the partially Euclidean nature of the domain.} We have proposed an alternative that considers the perspective of the covariance function. Using visibility graphs in the domain, we present visGP -- a method that respects domain geometry by encoding it into a graph of adjacency relationships between points and exploiting Dempster's method of covariance selection to simultaneously enforce marginal and conditional covariance (Markov) constraints. VisGP preserves stationary variances and Euclidean covariances on points that are connected via straight lines through the domain of interest. These properties, verified both theoretically and empirically, are unique to visGP among the competing methods, and they ensure that any analysis restricted to a convex subdomain of the non-convex domain coincides exactly with traditional GP analysis using Euclidean covariances. Computationally, we exploit chordal graphs to achieve a computationally efficient algorithm for visGP, devising a novel graph stochastic gradient descent algorithm. In all the simulations and the acidity level analysis, visGP performs well against state-of-the-art methods, consistently emerging as the best or competitive with the best. In terms of speed, it is the fastest algorithm. %
Future research %
will investigate the mathematical properties of parameter estimates in asymptotic regimes. We will also develop an open-access software for visGP for broader accessibility of the method.

\section*{Acknowledgments}
The authors are grateful for partial funding from the National Science Foundation (NSF) Division of Mathematical Sciences grant DMS-1915803 and for the use of the facilities at the Joint High Performance Computing Exchange (JHPCE) in the Department of Biostatistics, Johns Hopkins Bloomberg School of Public Health that have contributed to the research results reported within this paper.

\renewcommand\thesection{S\arabic{section}}
\renewcommand\theequation{S\arabic{equation}}
\renewcommand\thefigure{S\arabic{figure}}
\renewcommand\thetable{S\arabic{table}}

\setcounter{section}{0}
\setcounter{equation}{0}
\setcounter{figure}{0}
\setcounter{table}{0}

\clearpage
\section*{{\Large Supporting information for ``Visibility graph-based covariance functions for scalable spatial analysis in non-convex domains''}}\label{sec:non-convex_append}

\vskip 0.3in

\section{Detailed review of existing spatial analysis methods for non-convex domains}\label{web_app_A}
Multidimensional scaling (MDS) \citep{cox} is a very general approach that maps the locations from an arbitrary space, with some notion of a distance, into a Euclidean space while preserving inter-point distances as accurately as possible. For example, for original points $\{s_i\}\in \calD$ where $\calD$ denotes a non-Euclidean domain with pairwise distances $d_{ij}$, we might search for points $\{S^*_i\} \in \mathbb{R}^{k}$ for some $k$ with pairwise distances $\delta_{ij}$ which minimizes a loss function such as $\sum_{ij} \frac{ (\delta_{ij}- d_{ij})^2}{d_{ij}}$. However, this has a necessarily distorting effect, as it is not possible to preserve all the distances through the embedding. %
MDS is the theoretical basis of the popular algorithm ISOMAP \citep{tenenbaum}. ISOMAP maps from non-Euclidean geodesic distances to Euclidean distances. However, the geodesic distances are approximated by summed Euclidean distances of neighboring points, which would be an inaccurate procedure in our context unless the domain is nearly convex.

\cite{davis} propose to input intra-domain (i.e., geodesic) distances into any {\em Euclidean covariance function}, \blue{a valid (positive-definite) covariance function when used with Euclidean distances}, to construct a candidate covariance matrix, which is then passed through an algorithm to find the ``nearest" positive definite matrix to the candidate. Specifically, let $C([d_{ij}])$ denote the matrix of covariance values applying some covariance function to geodesic distances \blue{$d_{ij}$.} Then $C([d_{ij}])$ may not be positive definite, but it has eigen-decomposition $C([d_{ij}])=V \Lambda V'$. For some tolerance value $\epsilon > 0$, any negative entries of $\Lambda$ are replaced with $\epsilon$ to form a new non-negative diagonal matrix $\tilde{\lambda}$. Then $\tilde{C} =V \tilde{\Lambda} V'$ is taken as the new covariance matrix for kriging, since it is positive definite and hopefully close to $C$. However, the approximation could be poor if the original matrix has large negative eigenvalues. Additionally, the ascertainment of all pairwise geodesic distances and the eigen-decomposition of $C$ are computationally intensive for large data sets.

\cite{niu} present a method that uses the heat kernel as a covariance kernel; the heat kernel, in turn, is approximated by the transition probabilities of Brownian motion through the domain, where physical boundaries impede the motion of particles. However, these transition probabilities must be estimated by using simulations, which is computationally expensive for large datasets. \cite{dunson} present a related method {\em GLGP} that aims to alleviate the computational burden by appealing to the Graph Laplacian, which ``corresponds to the infinitesimal
generator of a random walk on the sampled data points." Specifically, for training locations $\{s_1, ..., s_m\}$ and test points $\{s_{m+1}, ..., s_{m+n}\}$, they define a kernel $k_\epsilon(s,s') = \exp(- \frac{ |S-S'|^2}{4\epsilon^2})$ and matrix $W_{ij} = \frac{k_\epsilon(s_i, s_j)}{q_\epsilon(s_i) q_\epsilon(s_j)}$ where $q_\epsilon(S) = \sum_{i=1}^{m+n} k_\epsilon(s, s_i)$. Then the ``graph" of interest is the complete, weighted graph over all points with weights $W$. Further defining the diagonal matrix $D_{ii} = \sum_{j=1}^{m+n} W_{ij}$, the Graph Laplacian is given by $L = \frac{D^{-1}W-I}{\epsilon^2}$, where $I$ is the identity matrix. The covariance function is then built using finitely many eigen-pairs of $L$. However, the approach does not respect domain boundaries as absolute (as reflected in the use of the Euclidean distance in defining the kernel); rather, it relies on a finely-tuned bandwidth parameter $\epsilon$, which can potentially lead to points close in Euclidean space but far in geodesic distance to unduly influence each other. Also, the corresponding likelihood is difficult to optimize, leading to a lack of scalability for large datasets. 

\cite{borovitskiy} generalize Euclidean kriging to the case of Riemannian manifolds without boundaries. They begin with the stochastic partial differential equation solution \citep{whittle} to the Mat\'{e}rn process given by  $
(\frac{2\nu}{\kappa^2}-\Delta)^{\nu/2+d/4} f = \mathcal{W}
$, where $f$ is the process and $\mathcal{W}$ is white noise, and the left-hand side contains Mat\'{e}rn parameters. This can be generalized to Riemannian manifolds by replacing the Laplacian $\Delta$ with the Laplace-Beltrami operator. However, this theory has no immediate application to the domains of interest here since we deal with manifolds \textit{with} irregular (sharp) boundaries (e.g., the shorelines). \blue{The Barrier Spatial Gaussian Field \citep{bakka} is a way of accounting for physical boundaries similarly using a stochastic partial differential equation model, but it requires a mesh approximation which can be computationally challenging}.

A recent notable contribution is BORA-GP \citep{jin2022spatial}, which encodes the geometry of the domain in the form of neighbor-relationships between points and proceeds by fitting a nearest neighbor Gaussian process \citep{vecchia, datta2016} in a Bayesian manner. Points are assumed to be conditionally independent of each other given the nearest-neighbor sets; this yields a local low-rank approximation of the likelihood. BORA-GP differs from the usual nearest neighbor approximation in that neighbor sets only include Euclidean neighbors, i.e., points connected by a straight line through the domain. This neighbor scheme is intuitive as it preserves the geometry of the domain and has similarities to the method we propose here but differs in that BORA-GP requires an ordering of the locations, which leads to a lack of stationarity for highly irregular domains. Also, BORA-GP does not attempt to preserve any covariance values relative to the Euclidean model, even though the Euclidean values are partly valid for \blue{partly Euclidean} domains.

\section{Proofs}\label{web_app_B}

\subsection{Proof of \blue{Proposition \ref{prop:marg}}}

\blue{We recall from Section \ref{sec:process} that} for any $s \notin \calV$, the neighbor sets $N(s)$ \blue{are subsets of $\calV$ that corresponds to cliques in the visibility graph $\calG$} and for any $s \in \calV$, $N(s)=\{s\}$. Then we can represent the process $w(\cdot)$ from (\ref{eq:finite}) and (\ref{eq:process}) as 
\begin{equation}\label{eq:ortho}
    w(s) = v(s)'w(\calV) + z(s) 
\end{equation}
where $v(s)$ is a vector encasing $B(s)$ and inserting zeros at locations in $\calV$ not corresponding to $N(s)$, and $z(s) \ind N(0,C(s,s)- C(s,N(s)) 
    C(N(s),N(s))^{-1}C(N(s),s))$ and $\{z(s) \given s \in \calD\} \perp w(\calV)$. Hence, we immediately have
\begin{align*}
    \blue{C^*(s,s)=}Var(w(s))&=v(s)'Var(w(\calV))v(s) + C(s,s)- C(s,N(s)) 
    C(N(s),N(s))^{-1}C(N(s),s)\\
    &=B(s)Var(w(N(s)))B(s)' + C(s,s)- C(s,N(s)) 
    C(N(s),N(s))^{-1}C(N(s),s)\\
    &=C(s,N(s))   C(N(s),N(s))^{-1}C^*(N(s),N(s))C(N(s),N(s))^{-1}C(N(s),s)\, + \\
    & \qquad C(s,s)- C(s,N(s)) 
    C(N(s),N(s))^{-1}C(N(s),s)).
\end{align*}

As the neighbor sets $N(s)$ are constructed to correspond to a clique (complete sub-graph) of $\calG$ (section \ref{sec:process}), by the property of covariance selection (\ref{eq:covsel}), the covariances from the original covariance $C$ are preserved on the cliques. Hence, we have $C^*(N(s),N(s))=C(N(s),N(s))$. This implies $C^*(s,s) = Var(w(s)) = C(s,s)$. As $C$ is isotropic, $C(s,s)=C(s',s')$ for all $s,s' \in \calD$, we immediately have $C^*(s,s)=C^*(s',s')$ for all $s,s' \in \calD$ and $w(\cdot) \blue{\sim GP(0,C^*)}$ is marginally stationary on $\calD$  proving (a). 

\subsection{\blue{Proof of Proposition \ref{prop:process}}}

\blue{We use the notations $\calV_n$, $\calG_n$, and $C^*_n$ to explicitly denote the dependence of these quantities on the sample size $n$.} For $s,s' \in \calV_n$ and connected in the domain, we exactly have $C^*_n(s,s')=C(s,s')$ for all $n$ from the properties of covariance selection. So the result is exact. 

We then consider $s,s' \notin \calV_n$ that are connected in $\calD$. \blue{Note that Equation (\ref{eq:ortho}) holds with $\calV=\calV_n$ for all $n$.} Then, as $z(s) \perp z(s')$ and both are independent of $w(\calV_n)$, we have
\begin{align}\label{eq:cov}
    C^*_n(s,s') =& Cov(v(s)'w(\calV_n),v(s')'w(\calV_n)) \nonumber \\
    =& C(s,N(s))   C(N(s),N(s))^{-1}C^*_n(N(s),N(s'))C(N(s'),N(s'))^{-1}C(N(s'),s')).
\end{align}

As $\calD$ is open and $s,s'$ are connected in $\calD$,  we can find open balls $O(s)$ and $O(s')$ around $s$ and $s'$, respectively, such that the balls lie entirely within 
 $\calD$ and each point in $O(s)$ is connected to each point in $O(s')$ in $\calD$. As $\cup_n \calV_n$ is dense in $\calD$, for large enough $n$, $N(s) \subset O(s)$ and $N(s') \subset O(s')$ implying that $$C^*_n(N(s),N(s'))=C(N(s),N(s')) \mbox{ for large enough } n $$ due to the property of covariance selection. Additionally, as $\cup_n \calV_n$ is dense in $\calD$, for every $s$, as $n$ increases each member of $N(s)$ converges to $s$, \blue{and the same applies for $N(s')$ and $s'$}. Hence, assuming $C(s,s)=1$ without loss of generality (as $C$ is isotropic), we have
 \begin{equation}\label{eq:lim1}
     \lim_n C^*_n(N(s),N(s'))= \lim_n  C(N(s),N(s')) = C(s,s')11' = C(s,s') \lim_n C(N(s),s)C(s',N(s')).
 \end{equation}
 
 Now consider a \blue{zero-mean} GP $u(\cdot)$ on $\calD$ equipped with the isotropic (Euclidean) covariance function $C$. Then
 $Var\big(u(s) \given u(N(s)\big) \leq Var\big(u(s) \given u(N(s)[1]\big)$ where $N(s)[1]$ denotes the first member of $N(s)$ implying
 \begin{equation}\label{eq:lim2}
     0 \leq \lim_n 1 - B(s)C(N(s),s) \leq \lim_n 1 - C(s,N(s)[1])^2 = 0, \mbox{ i.e., } \lim_n B(s)C(N(s),s) = 1.
 \end{equation}

Combining (\ref{eq:lim1}) and (\ref{eq:lim2}) in (\ref{eq:cov}), we prove part (b) as follows:
\begin{align*}\label{eq:cov2}   
    \lim_n C^*_n(s,s') =& C(s,s') \lim_n B(s)C(N(s),s)C(s',N(s'))B(s')' \\
    &+ \lim_n B(s) \Big[C(N(s),N(s')) -  C(s,s') C(N(s),s) C(N(s'),s') \Big]B(s')' \\
    =& C(s,s') \pm \lim_n \Big\|C(N(s),N(s')) -  C(s,s') C(N(s),s) C(N(s'),s') \Big\|_2\|B(s)\|_2\|_2B(s')\|_2 \\
    =& C(s,s') \pm 
    \lim_n o_p(1) M^2\\
    =& C(s,s'). 
\end{align*}

\blue{This proves the partially Euclidean property for points connected in the domain. To prove the Markovian property for points not connected in the domain,} we first consider $s,s' \in \blue{\cup} \calV_n$ \blue{that are not connected in $\calD$. Let $\calU_n = \{ s \in \calV_n \given s \in N(u) \mbox{ for some } u \in \calD \setminus \calV_n\}$. In other words, $\calU_n$ is the subset of $\calV_n$ consisting of points that are in the neighbor sets of at least one location outside $\calV_n$. As $V_n$'s are increasing with $\cup V_n$ being dense in $\calD$, and the neighbor sets are of size at most $k$, for any $s,s' \in \cup \calV_n$, there exists a positive integer $n(s,s')$ such that for all $n \geq n(s,s')$, $s,s' \in \calV_n$ and neither $s$ or $s'$ belong in neighbor sets for any location $u$ in $\calD$. For $n \geq n(s,s')$,} the $\sigma$-algebra generated by $\{w(u) \given u \in \calD \setminus \{s,s'\}\}$ is the same as the $\sigma$-algebra generated by $\{w(u) \given u \in \calV_n \setminus\{s,s'\}\} \cup \{ Z(u) \given u \in \calD \setminus \{s,s'\}\}$. It is easy to see that the former $\sigma$-algebra is generated by the latter, which follows directly from (\ref{eq:process}) \blue{as $\calU_n \subseteq \calV_n \setminus\{s,s'\}$ for $n \geq n(s,s')$}. The converse is \blue{also} true because $z(u)=0$ for all $u \in \calV_n$ \blue{and $s,s' \notin \calU_n$}. So, \blue{for $n \geq n(s,s')$,} we can write the conditional covariance 
 \begin{align*}
     \blue{C^*_n(s,s' \given \cdot) = }Cov\Big(w(s),w(s') &\given \big\{w(u) \given u \in \calD \setminus \{s,s'\}\big\}\Big)\\
     =& Cov\Big(w(s),w(s') \given \{w(u) \given u \in \calV_n \setminus \{s,s'\}\} \cup \{ Z(u) \given u \in \calD \setminus \{s,s'\}\} \Big) \\
     =& Cov\Big(w(s),w(s') \given \{w(u) \given u \in \calV_n \setminus \{s,s'\}\} \Big) \\
     =& (L^{-1})_{s,s'} \mbox{ where } L=C^*(\calV_n,\calV_n) \mbox{ from (\ref{eq:covsel})}\\
     =&0 \mbox{ as } s,s' \mbox{ are not connected in } \calD.
 \end{align*}
Here, we could drop all $z$ terms from the conditioning sets as $\{z(u)\}$ is a collection of independent random variables. This proves the Markov property (c) for all $s,s' \in \blue{\cup} \calV_n$ not connected in $\calD$ where the conditioning set is the $\sigma$-algebra generated by the entire process excluding the realizations at these two points. The result for the case where one of $s$ or $s'$ is not $\calV_n$ is true for any construction of the form (\ref{eq:process}) noting that as $\calV_n$ grows dense, the two points will have no neighbors in common.

\subsection{Proof of \blue{Proposition \ref{prop:convex_union}}}

Because there is a simple induction step to connect additional convex parts, it suffices to prove the result for a non-convex domain $\mathcal{D}$ can be decomposed into two smaller convex domains sharing one point in common, like the symbol for the number 8. Label one of the subdomains $\mathcal{A}$ and the other $\mathcal{B}$. Let $O$ denote the point that $\mathcal{A}$ and $\mathcal{B}$ have in common

Let $d_e(.,.)$ denote Euclidean distance and $\blue{d_{geo}}(.,.)$ denote geodesic distance. 

For $d \in \mathcal{D}$, let $|d| = d_e(d, O)$. Then for $d_1, d_2 \in \mathcal{D}$, 

\[ \blue{d_{geo}}(d_1, d_2) = \begin{cases} 
      d_e(d_1, d_2) & d_1, d_2 \in \mathcal{A} \quad or \quad d_1, d_2 \in \mathcal{B} \\
      |d_1| + |d_2| & else \\
   \end{cases}
\]

Let $\mathbf{A} = a_1, a_2, ..., a_n \in \mathcal{A}$ and $\mathbf{B} = b_1, b_2, ..., b_m \in \mathcal{B}$. Take $\blue{\calV}=(a_1, ..., a_n, O, b_1, ..., b_m)$.

Let $C_e$ denote the exponential covariance function with Euclidean distance and $C_w$ denote the exponential covariance function with water distance, \blue{both with parameters $(\phi, \sigma^2)$.}

Let $C^A = \blue{(C_e(a_1,O),\ldots,C_e(a_n,O))'}$ %
and define $C^B$ similarly.
Let $C^\dagger(s_i, s_j) = C_e(s_i, s_j) -\sigma^{-2}  C_e(s_i, O) C_e(s_j, O)$ \blue{denote the conditional GP covariance.}

\blue{We define a GP $\omega$ on $\calV$ as:} 
\begin{equation}\label{eq:union}
\begin{pmatrix} \omega(\mathbf{A}) \\ \omega(O) \\ \omega(\mathbf{B})
\end{pmatrix} \,{\buildrel d \over =}\, \begin{pmatrix} \sigma^{-2} \omega(O) C^A +\mathbf{z_1}  \\ \blue{N(0,\sigma^2)} \\  \sigma^{-2} \omega(O) C^B +\mathbf{z_2}  \end{pmatrix}
\end{equation}
where $\mathbf{z_1}, \mathbf{z_2}, \omega(O)$ are mutually independent, $\mathbf{z_1} \sim GP(0, C^\dagger(\mathbf{A}))$, and $\mathbf{z_2} \sim GP(0, C^\dagger(\mathbf{B}))$. 
We will show that the right-hand-side expression \blue{corresponds to a visGP on $\calV$ and the visGP covariance function equals $C_w$. First note that}  it is clear that $\omega(\mathbf{A}) \perp \omega(\mathbf{B}) | \omega(O)$, and a pair of points can only be disconnected if one belongs to $\mathbf{A}$ while the other belongs to $\mathbf{B}$. \blue{This proves the Markov property required for visGP.}

\blue{We now establish the marginal stationarity and partially Euclidean property of the distribution on the right side of (\ref{eq:union}).} %
The $i,j$ entry of $\Var (  \sigma^{-2} \omega(O) C^A +\mathbf{z_1})$ is given by
\[
\sigma^{-4} \Cov(C_e(a_i, O)\omega(O),  C_e(a_j, O) \omega(O)) + C^\dagger(a_i, a_j)\]
\[
= \sigma^{-2} C_e(a_i, O) C_e(a_j, O) + C_e(a_i, a_j) -\sigma^{-2} C_e(a_i, O)  C_e(a_j, O)  
\]
\[
= C_e(a_i, a_j) = C_w(a_i, a_j).
\]

\blue{Here the last equality holds as the Euclidean and geodesic distances are the same within $\mathcal A$.} 
So, we have $\Cov(\omega(\mathbf{A}))  \blue{= (C_e(a_i,a_j))= (C_w(a_i,a_j))}$. Similarly, $\Cov(\omega(\mathbf{B}) ) \blue{= (C_e(b_i,b_j))= (C_w(b_i,b_j))}$.

\blue{Marginal stationarity is then immediate as for any $a_i \in \mathbf A$, we have $\Var(\omega(a_i))=C_e(a_i,a_i)=\sigma^2$, and same for any $b_i \in \mathbf B$, and $\Var(\omega(O))=\sigma^2$ from \eqref{eq:union}.}

Also, $
\Cov(\omega(O), \omega(a_i)) = \Cov(\omega(O),\sigma^{-2} C_e(O, a_i) \omega(O) + \mathbf{z}_{1i})
= C_e(O, a_i) = \blue{C_w(O, a_i)}$ \blue{as $a_i,O \in \mathcal A$. Similarly, $
\Cov(\omega(O), \omega(b_i)) = C_e(O, b_i) = \blue{C_w(O, b_i)}$.}

Finally, \blue{for $a_i \in \mathbf A$ and $b_j \in \mathbf B$, noting that $d_{geo}(a_i,b_j)=|a_i| + |b_j|$, we have}
\begin{align*}\label{eq:covunion}
\blue{\Cov(\omega(a_i),\omega(b_j)) =}& \Cov(\sigma^{-2} C_e(O, a_i) \omega(O) + \mathbf{z}_{1i}, \sigma^{-2}C_e(O, b_j) \omega(O) + \mathbf{z}_{2j})\\
=& C_e(O, a_i)C_e(O, b_j)\sigma^{-4}\Cov( \omega(O) ,  \omega(O) )\\
=& C_e(O, a_i)C_e(O, b_j)/\sigma^2\\
=& \{\sigma^2\exp(-\phi |a_i|)\}\{\sigma^2 \exp(-\phi |b_j|)\} /\sigma^2 \\
=& \sigma^2 \exp(-\phi ( |a_i| + |b_j|))\\
=& C_w(a_i, b_j).
\end{align*}

This completes the proof that $\Var(\omega(\calV))=C_w(\calV,\calV)$, i.e., the covariance between the GP (\ref{eq:union}) at any two locations of $\calV$ is given by the exponential covariance using water distances. 

\blue{For $\omega(\calV)$, we have already proved marginal stationarity, being Markov on points not connected in the domain, and the partial Euclidean covariance property within $\mathbf A \cup \{O\}$ or within $\mathbf B \cup \{O\}$. It only remains to show that the partial Euclidean covariances also hold for any $a_i \in \mathbf A$ and $b_j \in \mathbf B$ which is connected in the domain. For that to hold, as $\mathcal A \cap \mathcal B = \{O\}$, the point $O$ must lie on the straight line between $a_i$ to $b_j$. Then, $|a_i- b_j| = |a_i| + |b_j|$ and we immediately have $Cov(\omega(a_i),\omega(b_j)) =\sigma^2 \exp(-\phi ( |a_i| + |b_j|)) = \sigma^2 \exp(-\phi ( |a_i- b_j|)) = C_e (a_i,b_j)$. Thus, the GP defined in (\ref{eq:union}) satisfies all three properties (marginal stationarity, partially Euclidean, and Markov). By the property of covariance selection, these three properties are unique to the visGP covariance matrix, hence $\omega(\calV)$ is indeed the visGP.}

\section{Algorithms}\label{web_app_C}
\blue{In this section, we present implementation details of the methods used in the simulation studies. We fit four classes of models. A nearest-neighbor Gaussian process using Euclidean distances was fit using the \texttt{BRISC} package in \texttt{R} \citep{saha,brisc}, with 10 neighbors and an exponential covariance \citep{saha}. \blue{The GLGP model \citep{dunson} uses a grid search on the number of eigenvectors in $\{50,100,150\}$ and bandwidth in $\{\exp(r): r = -2, -1.8, -1.6, \dots, 2 \}$ with numerical optimization over the error variance and diffusion time.}
 Due to computational constraints, the GLGP results are omitted for the medium- and large-sample analyses, and interval estimates are not calculated for GLGP. BORA-GP \citep{jin2022spatial} is fit with diffuse priors, exponential covariance, 10 nearest neighbors, and an ordering based on the first coordinate of $s$; if that ordering fails, the ordering by the second coordinate of $s$ is used. \blue{There are $10,000$ posterior samples with $5,000$ discarded for burn-in}. %
Finally, we fit the visGP model with exponential covariance and carry out (with $10$ nearest neighbors) the three distinct prediction strategies described in detail later in Section \ref{web_app_C}. The full likelihood function is used for the $n=250$ and $n=1,200$ scenarios, but the graph stochastic gradient descent strategy described in Section \ref{sec:sgd} is used for the $n=10,000$ scenario with $5,000$ clique iterations and a distance threshold of 1 unit.}

\subsection{Details of the graph stochastic gradient descent algorithm}\label{sec:sup_sgd}
We use a version of stochastic gradient descent called ``Root Mean Square Propagation" or ``RMSProp" which is designed to improve convergence by using an exponentially weighted moving-average gradient \citep{goodfellow}. We describe this algorithm below.

\begin{algorithm}[ht]{\linewidth - 0.5 in}
	\caption{Stochastic gradient descent for Gaussian likelihood maximization}\label{alg:sgd}
	\begin{algorithmic}
		\State Set learning rate $\alpha$, decay rate $\beta$, small stability constant $\epsilon$, maximum number of iterations $T$, and initial parameter estimate $\hat\theta$
		\State Initialize accumulation variables $v= \mathbf{0}$
		\For{$t=1$ to $T$}
		\State Randomize the clique and separator sequence corresponding to the perfect ordering
		\For{i = 1 to number of cliques}
		\State For the $i^{th}$ clique $K_i$ and the $i^{th}$ separator $S_i$, compute gradient of log-likelihood:
		\State $g \gets \nabla_{{\theta}} \log L^{(i)} = \nabla_{{\theta}}[\log f(Y_{K_i})- \log f(Y_{S_i})]$
		\State Update accumulation variables:
		\State $v \gets \beta*v+(1-\beta)g*g$
		\State Compute step size and update parameter estimates:
		\State ${\theta} \gets {\theta} + \alpha g / \sqrt{v+\epsilon}$
		\EndFor
		\EndFor
	\end{algorithmic}
\end{algorithm}

\newpage
\subsection{Nearest neighbor clique likelihood}\label{sec:nngp}

The SGD helps improve scalability when the number of cliques in the perfect ordering is large. However, evaluating each clique (or separator) likelihood requires computation time that is cubic in the size of the clique, which will be prohibitive if the size of some cliques themselves are large. Note that a clique likelihood $N(Y(K) \given X(K)'\beta, C(K,K) + \tau^2 I)$ is simply a standard GP likelihood with a Euclidean covariance over a set of points that are fully connected in the domain $\calD$ (i.e., the clique lies in a convex subdomain of $\calD$). In Euclidean domains, local low-rank approximations like the nearest neighbor GP \citep{datta2016, datta} which assume independence of responses conditional on nearby neighboring sets of points, offer excellent linear-time approximations to the full GP likelihood within each clique. Note that the neighbor sets used within each clique to create an NNGP approximation to the clique likelihood are different from the neighbor sets created using the visibility graph and used to define the visGP process. 

\subsection{Distance thresholding}\label{sec:dist}

Above, we defined $\calG = (\calV,E)$ to be the adjacency graph, where connections indicated whether the line segment connecting a pair of points lay entirely within the non-convex domain. For large datasets, one can amend this to a {\em distance-thresholded visibility graph}, where two points are adjacent if and only if the corresponding line segment lies in the domain \textit{and} the distance between the points is bounded by some user-defined threshold distance $d_{max}$. This introduces sparsity into the adjacency matrix and can be expected not to significantly skew results as long as $d_{max}$-balls around observed points typically contain a fair number of other observations. In other words, as long as there are many neighbors within the threshold, the threshold itself may not be crucial as dependence still flows through the neighbor connections.

\subsection{Prediction strategies}\label{sec:prediction}

In Section \ref{sec:process}, we demonstrated how the visGP can be extended from a finite set $\calV$ to a process on the entire domain $\calD$. The construction imposes minimal restrictions on the choice of the neighbor sets, except that the neighbors of a location need to be connected through the domain to the location and to each other. This is because, if two neighbors are not connected through the domain, there is no clear distance we should associate with the pair, \blue{which is needed to compute their covariance, and subsequently the nearest-neighbor kriging weights $B(s)$ in (\ref{eq:process}).} If the Euclidean distance is used, the geometry of the domain is not respected; if the through-domain distance is used, the corresponding covariance matrix may not be positive definite. Note that this issue was not addressed in \cite{jin2022spatial}, who assume the Euclidean covariance among all locations in the conditioning set, although they may not be connected through the domain when defining the nearest-neighbor kriging weights. 

Selecting neighbor sets that are mutually connected through the domain is critical to visGP possessing the desirable properties, as established in \blue{Propositions \ref{prop:marg} and \ref{prop:process}.} 
We propose \blue{two} different algorithms \blue{({\em nearest clique} and {\em maximum precision})} for choosing \blue{neighbor sets for visGP. Each strategy for choosing neighbor sets leads to a corresponding prediction strategy using (\ref{eq:process}). Additionally, we consider a {\em precision-weighted} prediction strategy that averages over predictions from different choices of neighbor sets. We detail all three strategies below. Note that in each of these three strategies, if there are only $m$ adjacent observations or candidate neighbors for the new location, where $m$ is less than the prescribed number of neighbors $k$, then just those $m$ points will be used in the prediction.}

\begin{enumerate}[labelwidth=2cm]
    \item Nearest clique (NC): In sequence, add the first nearest neighbor, the second nearest neighbor, and so on, until adding the next nearest neighbor would form an incomplete sub-graph. Specifically, let $N^m (u_i)$ denote the $m^{th}$ nearest neighbor of $u_i$, where candidate neighbors only include points connected to $u_i$ through the domain, so for an ordinary $k$-nearest neighbors scheme we would have $N(u_i) = \{N^1 (u_i), N^2(u_i) , ..., N^k(u_i)\}$. For an ordered set of locations $A$, let $\mathbf H(A)=1$ if all locations in $A$ are mutually connected, and $0$ otherwise. Then we follow the algorithm below:

    \begin{algorithm}{\linewidth - 0.5 in}
    \caption{Nearest clique algorithm}
    \begin{algorithmic}
\State $N \gets \emptyset$
\For{i = 1, 2, ..., k} 
\If{$\mathbf H( N \cup N^{i}(u_i) ) = 1$}
$N \gets  N \cup N^{i}(u)$
\Else

\textbf{break}
\EndIf
\EndFor

\Return $N$
\end{algorithmic}
\end{algorithm}

    \item Maximum precision (MP): Of all maximal cliques of the graph with $k$ nearest neighbors, we choose the clique whose conditional predictive variance is smallest. Specifically, let $\mathcal{Q}$ denote the set of maximal cliques of the visibility graph associated with $N(u_i)$. Then
    
\begin{minipage}{\linewidth - 0.5 in}
    \[
    N_{MP}(u_i) = \argmin_{Q \in \mathcal{Q}} \mathbf{C}_{u_i, u_i}- \mathbf{C}_{u_i, Q} \mathbf{C}^{-1}_{Q, Q} \mathbf{C}_{Q, u_i}
    \]
\end{minipage}

\vspace{1em}
  
    \item Precision-weighted (PW): 
    Instead of considering a single neighbor set, we consider a series of neighbor sets from non-overlapping cliques, starting with the largest nearest clique, then finding the next largest clique, and so on. The cliques are constrained to be non-overlapping by deleting them from the graph after they are selected, before beginning the search for the next clique. We calculate the kriging prediction for each clique and take the average of the predictions weighted by their conditional precision. Specifically, for a set of locations $A$, let $\mathbf{L}(A)$ return the largest clique that can be created from the members of $A$. Then
\[
    N_{PW}^{(1)}(u_i) = \mathbf{L}(N(u_i))
    \]
    \[
    N_{PW}^{(m)}(u_i) = \mathbf{L} (N(u_i) \setminus \cup_{p=1}^{m-1} N_{PW}^{(m)}(u_i) )
    \]
\[K^{(m)} = [\mathbf{C}_{u_i, u_i}- \mathbf{C}_{u_i, N_{PW}^{(m)}(u_i)} \mathbf{C}^{-1}_{ N_{PW}^{(m)}(u_i),  N_{PW}^{(m)}(u_i)} \mathbf{C}_{ N_{PW}^{(m)}(u_i), u_i}]^{-1}\]
\[
 \mu^{(m)} = \mathbf{C}_{u_i, N_{PW}^{(m)}(u_i)} \mathbf{C}^{-1}_{N_{PW}^{(m)}(u_i), N_{PW}^{(m)}(u_i)} \omega_{N_{PW}^{(m)}(u_i)}
\]
\[
E( w(u_i) \given \calV) = \frac{\sum_p K^{(m)} \mu^{(m)}}{\sum_p K^{(m)}}.
\]

\end{enumerate}

{\color{black}
\section{More on chordal completion}\label{web_app_D}
One may be concerned that the strategy of adding edges to the visibility graph to create a chordal graph introduces arbitrary distortions in the effective geometry of the domain. We have found heuristically that the distortion will tend to be minimal.

For example, we investigated the edges added for 500 samples of the fork-shaped domain where $n=200$, which yields $19,900$ total possible edges. Below, we show the frequency of how many edges are added to achieve chordality.

\includegraphics[scale=.5]{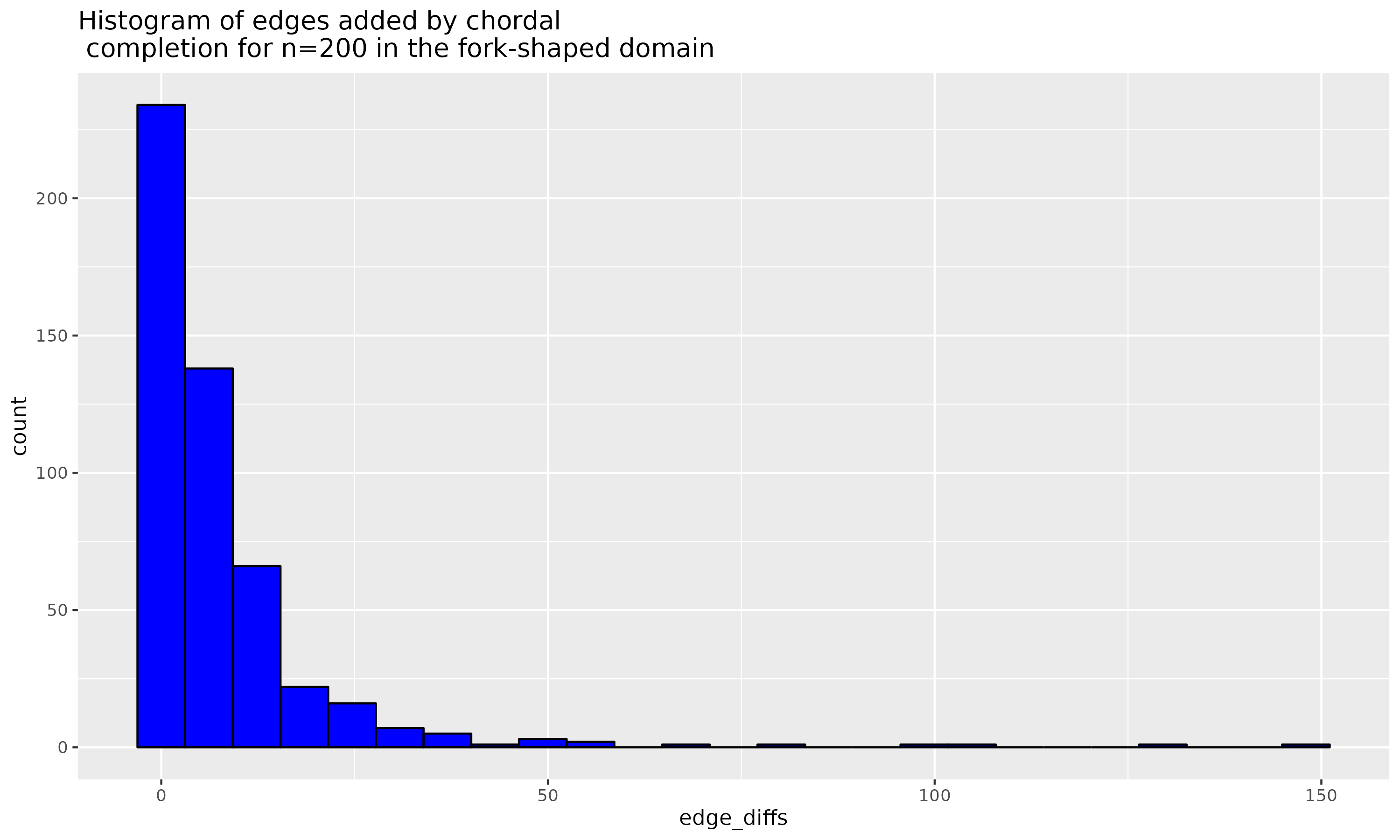}

Clearly, the number of edges added tends to be very small (mostly $<50$) relative to the total number of possible edges ($19,900$). Furthermore, we consider the instance (among the $500$ runs) where the largest number of edges (148) had to be added and plot the added edges shown in green below.

\includegraphics[scale=.5]{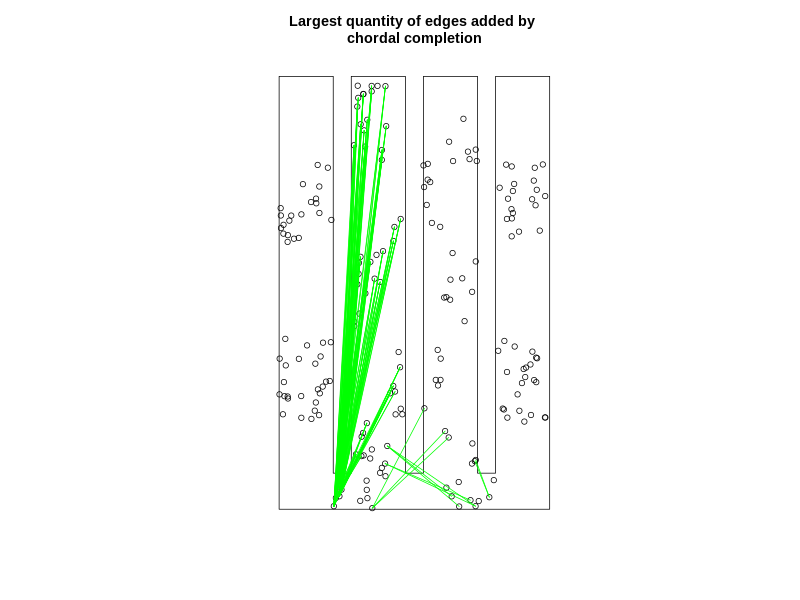}

The majority of edges connect to one point, and they are edges that ``hug" or ``clip'' the boundary, rather than cutting across branches of the domain. This means that the Euclidean distance is close to the geodesic distance, and thus, considering them adjacent does not significantly distort the geometry.

To understand this intuitively, note that the only situation where chordal completion is problematic is when it joins a pair of points whose Euclidean distance is small but whose geodesic distance is large. Such a join is unlikely as it will tend to create many new cycles, which will require new edges. %
Formally, if two points $A$ and $B$ are close in Euclidean distance but far apart in geodesic distances, it is likely that there will be several other pairs of points $A'$, $B'$ in the domain such that $(A,A')$, $(A',B')$ and $(B',B)$ are connected. Joining $A$ and $B$ would then lead to several new $4$-cycles of the form $A,A',B',B,A$, requiring the addition of many other edges to make each such cycle chordal. Minimal or near-minimal chordal completions will thus try to avoid such a join. This is illustrated in the schematic below, which shows the area for choosing such points $A'$ and $B'$ is much larger when $A$ and $B$ are close in Euclidean distance but far apart in geodesic distance. 

\begin{figure}[!ht]
    \centering
    \includegraphics[trim={200 1600 0 300}, clip, scale=.25]{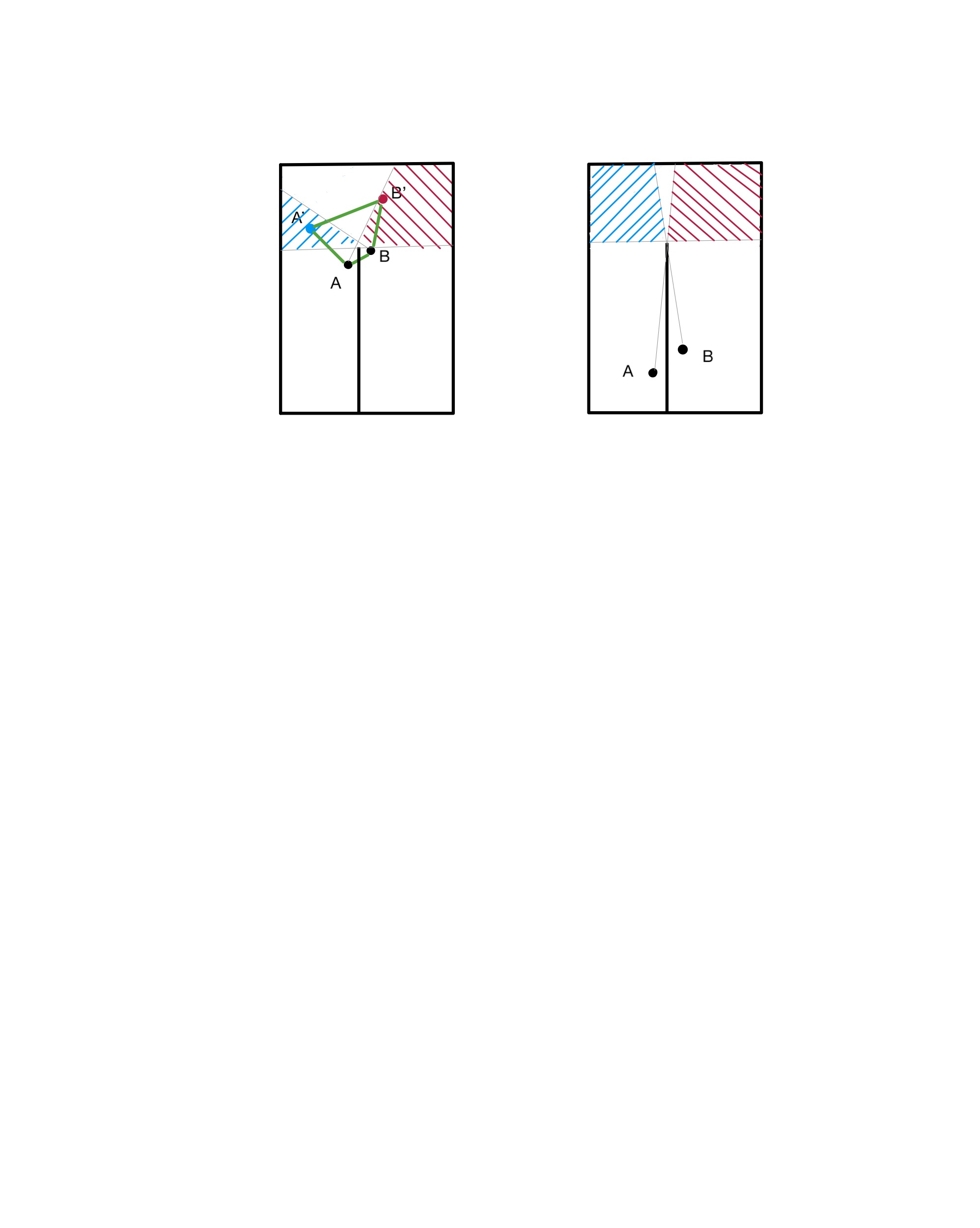}
    \caption*{\blue{Figure: Illustration of why chordal completion is less likely to join points that are close in Euclidean distance but far apart in geodesic distance. We consider a U-shaped domain akin to Figure \ref{fig:covcomp_locs}. The left figure considers two points $A$ and $B$ that are close in both Euclidean and geodesic distances. If $(A,B)$ is joined by an edge, for each choice of $A'$ in the blue-shaded region and $B'$ in the red-shaded region, $A,A',B',B, A$ is a 4-cycle without any chords (example choices of $A'$ and $B'$ are illustrated). The right figures consider the same domain but different points $A$ and $B$ that are close in Euclidean distance but much further apart in geodesic distance. Now, the blue- and red-shaded regions are much larger. This shows that if $(A,B)$ are joined in the right figure, it would lead to more 4-cycles and consequently more addition of edges to make these cycles chordal than if $(A,B)$ are joined in the left figure.}}
    \label{fig:enter-label}
\end{figure}

Of course, this intuition may fail for more complicated (and, perhaps, pathologically) non-convex geometries, in which case making plots to study which edges are being added is warranted to ensure that a large number of undesirable edges are not being added due to the chordal completion.

}

\section{Details of the simulation study}\label{web_app_E}

The fixed function $f$ for the simulations in the fork-shaped domain is calculated on $20,000$ points, which are sampled for each run of the simulations. \blue{This finite set of points was resampled in every replicate simulation so that the adjacency relations did not have to be calculated for every simulation run. That is, we had $20,000$ distinct points and took different subsamples for each simulation run.}
Using the source points $p_1 = (-5, -5), p_2 = (-3, -5), p_3 = (-1, -5), p_4=(1,-5)$, which lie at the base of each prong of the fork, we create a function $f$ over the domain as follows:

\vspace{1em}

\begin{minipage}{\linewidth - 0.5 in}
\[
d_i(s) = d_g(p_i, s );\; i=1,...,4,\,
f^*(s) = d_1^2/3 + 3*\sin(d_3) - d_2*d_4,
f = \frac{f^* - mean(f^*)}{sd(f^*)}
\]
\end{minipage}

\vspace{1em}

where $d_g$ is the distance metric calculated through the domain, as above. 

\begin{figure}[!ht]
    \centering
    \begin{subfigure}[t]{.49\textwidth}
        \centering
        \includegraphics[scale=.383]{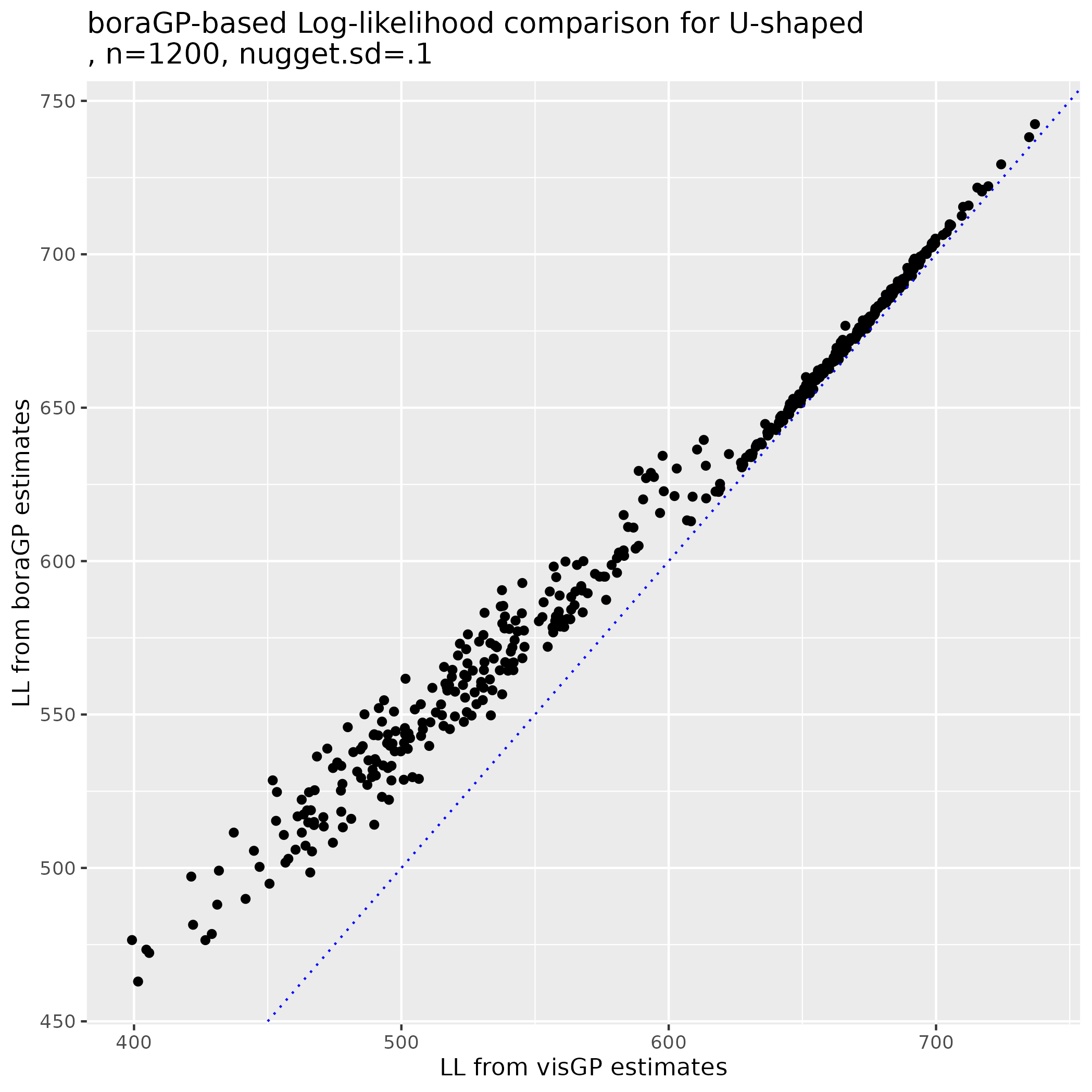}
        \caption{\blue{Log-likelihood comparisons under BORA-GP model}}
    \end{subfigure}
    \begin{subfigure}[t]{.49\textwidth}
        \centering
        \includegraphics[scale=.383]{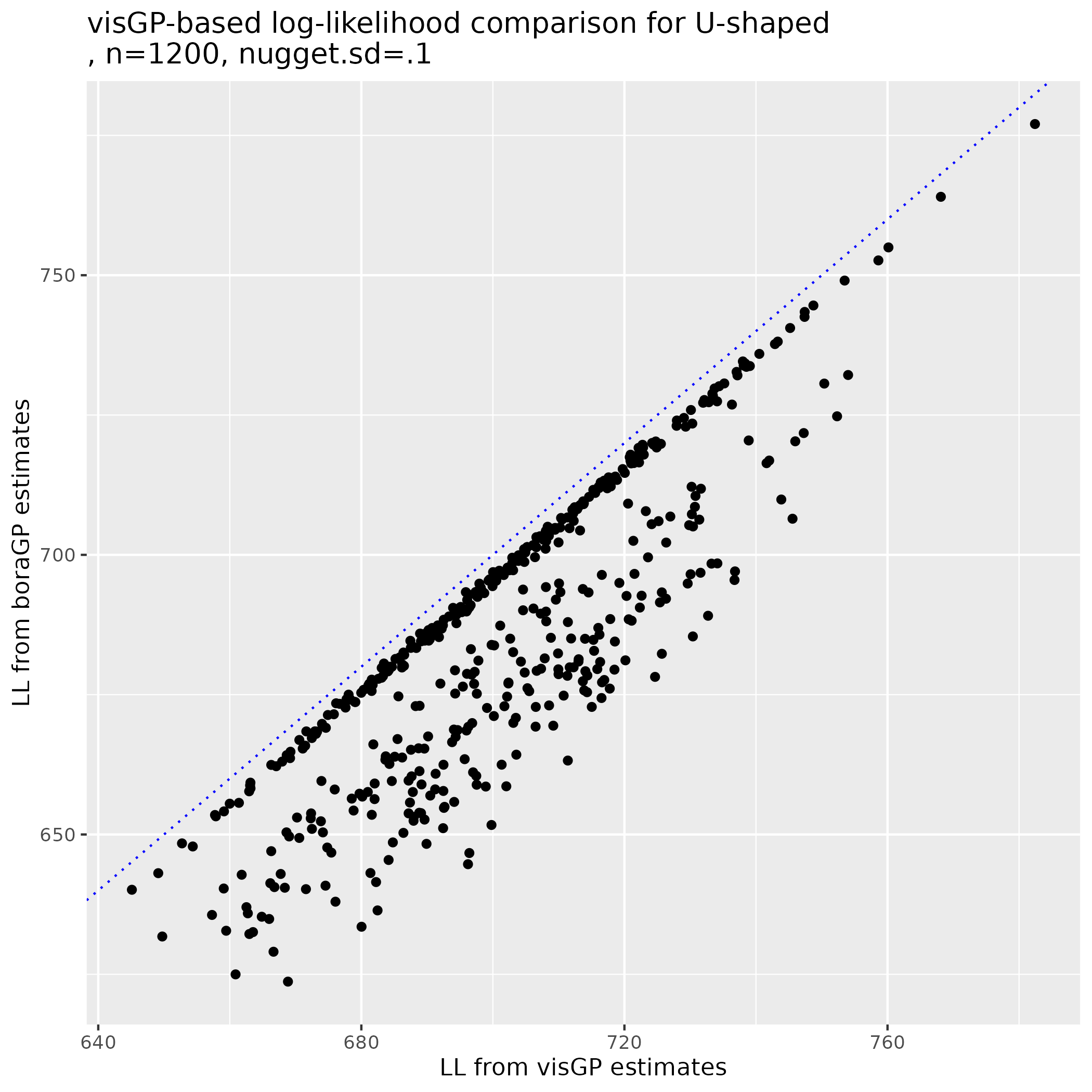}
        \caption{\blue{Log-likelihood comparisons under visGP model}}
    \end{subfigure}\\
    \begin{subfigure}[t]{.49\textwidth}
        \centering
        \includegraphics[scale=.383]{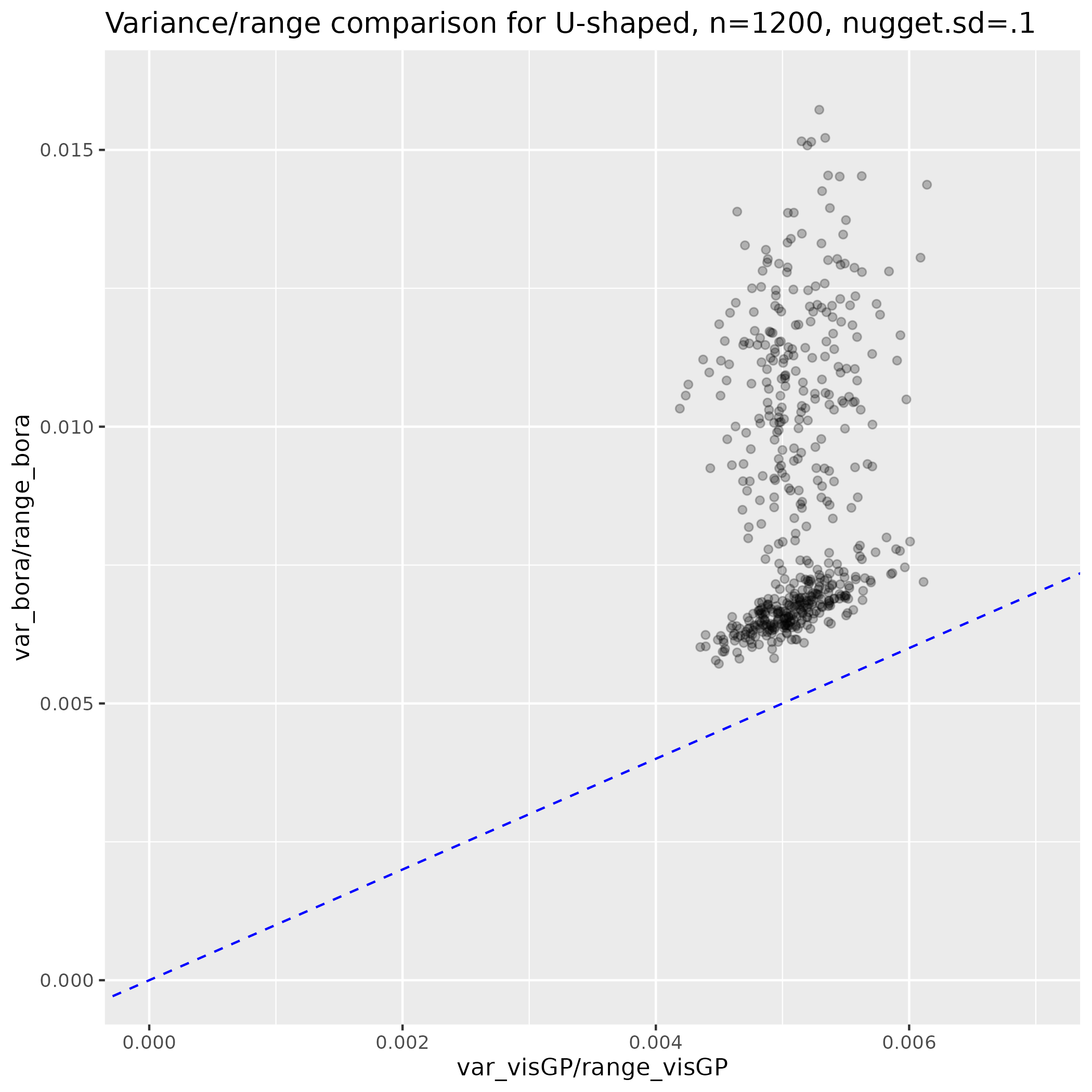}
        \caption{\blue{Ratio (variance/range) comparison}}
    \end{subfigure}
    \caption{\blue{Comparison of BORA-GP and visGP parameter estimates in the U-shaped domain. Parameters compared are MLE for visGP and posterior medians for BORA-GP.}}\label{fig:appf}
\end{figure}

\blue{
\section{Parameter Comparison of BORA-GP and visGP for U-shaped domain}\label{web_app_F}
In order to investigate the sources of discrepancy between the confidence interval lengths of BORA-GP and visGP, we examined their parameter estimates in the U-shaped domain shown in Figure 2a of the main text. For the following results, we used $n=1200$ and a nugget variance value of $0.01$.
The corners of the domain are the points $(\pm 6, \pm 6)$, and the central joint is at $Z=(0, 2)$. The mean process value at each point $s$ was generated as $d_g(Z,s)^3 + \sin(3*d_g(Z, s))$, scaled and normalized, where $d_g$ is the geodesic distance. 
In Figure \ref{fig:appf}, we compare the point (maximum likelihood) estimates from visGP with posterior median estimates from BORA-GP. Figures (a) and (b) compare the log-likelihoods arising from each model under each model's estimated parameters. These plots suggest that, as one would expect, each model tends to better optimize its own likelihood (compared to the other model), although the log-likelihood values are often similar. Figure (c) compares the ratio of spatial variance to range for each model; a higher value of this ratio is associated with higher predictive uncertainty. Although there is a clear correlation across the models, the fact that visGP tends to have a lower value for this metric may explain why its confidence intervals tend to be shorter.
}

\section{\blue{Implementation details for spatial analysis of pH levels in Chesapeake Bay}}\label{sec:phdetails}

\begin{figure}[!ht]
\centering
    \includegraphics[scale=.54]{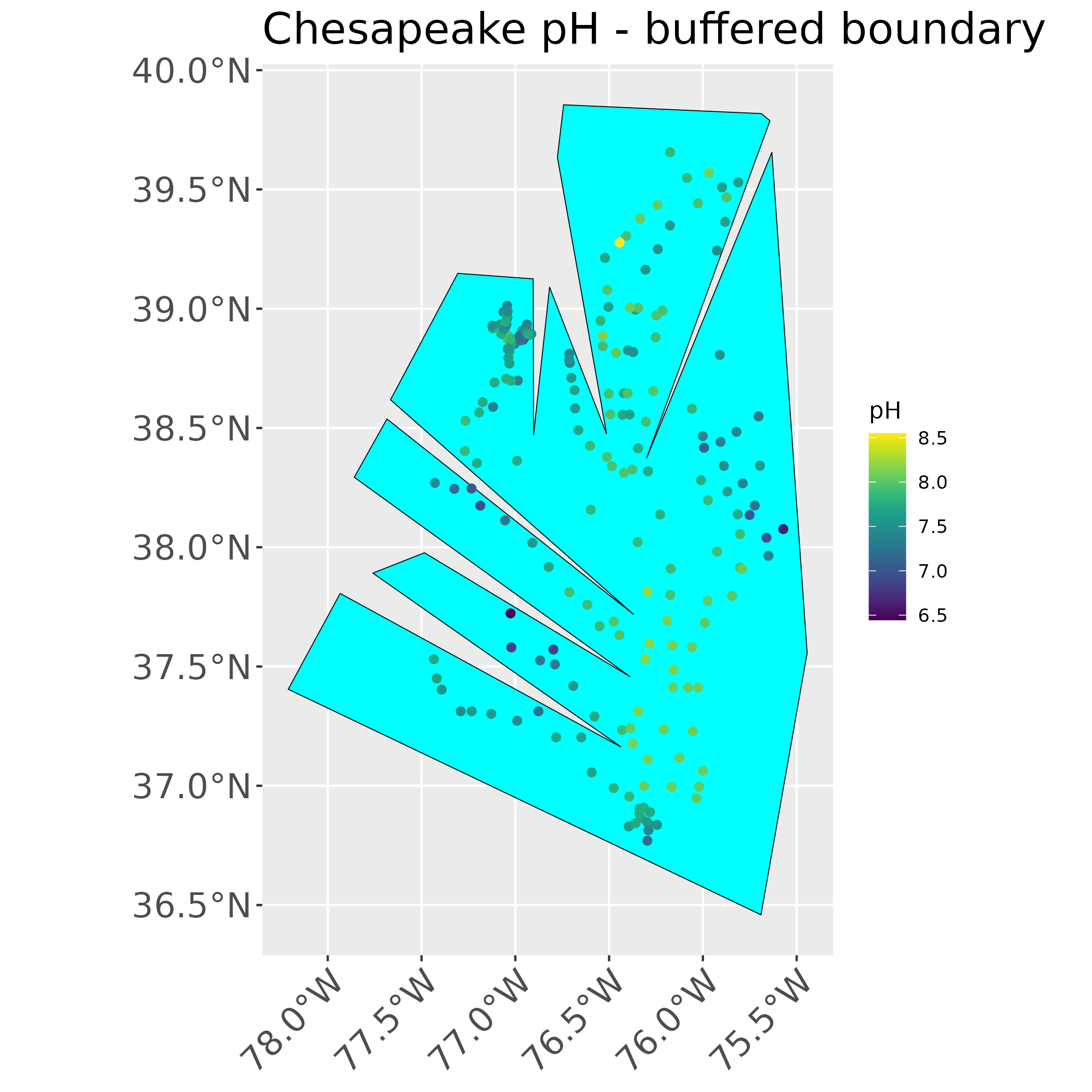}
    \caption{Buffered Chesapeake domain with average pH levels.}
    \label{fig:pH_buff}
\end{figure}

For the non-Euclidean methods, we face an issue that the monitoring stations are so close to the shoreline (in fact, some stations appear to be inland relative to the specified boundary file) that many lack any connections to other points which are connected strictly in the water. To address this, we ``buffer" the Bay's boundary while respecting the broad contours of the Bay's geometry, as can be seen in Figure \ref{fig:pH_buff}.

\blue{
Also, although we have been focused on non-Euclidean distances arising from non-convex domains, it is important to recognize that in this domain, all distances are non-Euclidean due to the curvature of the earth. To address this, we calculate actual geographic distances between points with the \texttt{geosphere} R package \citep{hijmans2017package} and get suitable points in $\mathbb{R}^2$ by multidimensional scaling. (This does not introduce any appreciable error in the interpoint distances because the region is small in area.) These locations are normalized and fed into the models. For visGP, we can calculate the adjacency relations based on actual geodesic segments. For BORA-GP, the adjacencies are based on the Euclidean relations of the raw latitude/longitude values since the relevant \texttt{barrier\_neighbor} function is internal to the \texttt{BORA-GP} package.}

\section{\blue{Details of the simulation study on process properties}}\label{web_app_H}

\blue{We present details of the data generation and competing methods used for the simulation study of Section \ref{sec:process}. %
We consider locations in a $U$-shaped domain (Figure \ref{fig:covcomp_locs}) with side lengths of 12 units 
and use a parent Euclidean covariance based on the Mat\'{e}rn function with spatial variance $\sigma^2=1$, smoothness $\nu =1$, inverse-range $\phi = 0.1$, and nugget variance $\tau^2=1$. %

For these comparisons, we use a modification of the BORA-GP algorithm which sets the grid size for proxy neighbor sources to $0.01$ \cite[see Figure 4,][]{jin2022spatial} rather than the width of the barrier crossing, since the barrier has $0$ width.
BORA-GP relies on an ordering of the locations, an inherent aspect of NNGP-type constructions. 
The effect of ordering on the variance of NNGP has been observed in convex domains \citep{Datta2021} and can be partly mitigated by the use of random orderings, where the decrease in variance does not get confined to one part of the domain \citep{guinness}. However, for BORA-GP in non-convex domains, the impact of ordering is exacerbated because standard orderings (e.g., $x$- or $y$-coordinate) will lead to systematic discrepancies in modeled variance in different parts of the domain, as we observe in Figure \ref{fig:covcomp_ord}. A random ordering also cannot be used as it will lead to many points without any neighbors. This is because random ordering leads to the selection of many distant points in the neighbor set, and in a non-convex domain, there will be likely barriers between these points precluding their inclusion into the neighbor set.

We also considered the possibility of comparing a method like that of \cite{davis} for this study. However, for this particular choice of domain and parameter values, the geodesic-distance-based matrix indeed turns out to be positive definite. So, methods like that of \cite{davis} are redundant as they will leave the Mat\`ern covariance on geodesic distances unchanged. However, in general, such positive definiteness is not guaranteed for arbitrary non-convex domains and parameter combinations. Applying the method of \cite{davis} would entail a positive-definiteness check for each update of the parameter values and, if needed, a projection of the geodesic-distance-based covariance matrix into the cone of positive-definite matrices. Such a projection is guaranteed to destroy the properties of marginal stationarity and partially Euclidean covariances. Also, the projection involves obtaining the singular-value decomposition of the matrix, typically requiring $O(n^3)$ computation for every iteration during optimization or sampling, and is thus not feasible for large $n$. Instead, we compared the covariances induced by a more pragmatic method, applying multi-dimensional scaling (MDS) on the geodesic-distance matrix to obtain a Euclidean embedding of the locations in $\mathbb{R}^3$ and applying Euclidean GP covariances.}

\newpage

\section{Simulation results in the fork-shaped domain}\label{web_tab_1}
\begin{longtable}{p{0.05\linewidth} | p{0.045\linewidth} | p{0.315\linewidth} | p{0.165\linewidth}|p{0.045\linewidth} |p{0.165\linewidth} }
\centering
n & $\sigma_{nug}$ & Method & MSE & CP & CI length \\
  \hline
$250$ & $0.1$ & BORA-GP & $5.450 \times 10^{-2}$ & 98\% & $9.806 \times 10^{-1}$ \\ 
  $250$ & $0.1$ & visGP: Maximum precision & $4.084 \times 10^{-2}$ & 92\% & $6.133 \times 10^{-1}$ \\ 
  $250$ & $0.1$ & visGP: Nearest clique & $4.160 \times 10^{-2}$ & 92\% & $6.176 \times 10^{-1}$ \\ 
  $250$ & $0.1$ & visGP: Precision-weighted & $4.046 \times 10^{-2}$ & 92\% & $5.974 \times 10^{-1}$ \\ 
  $250$ & $0.1$ & visGP: Standard kriging & $4.045 \times 10^{-2}$ & 92\% & $6.105 \times 10^{-1}$ \\ 
  $250$ & $0.1$ & Euclidean & $8.833 \times 10^{-1}$ & 75\% & $1.259 \times 10^{0}$ \\ 
  $250$ & $0.1$ & GLGP & $8.784 \times 10^{-1}$ &  &  \\ 
  $250$ & $0.25$ & BORA-GP & $1.464 \times 10^{-1}$ & 97\% & $1.548 \times 10^{0}$ \\ 
  $250$ & $0.25$ & visGP: Maximum precision & $1.108 \times 10^{-1}$ & 94\% & $1.203 \times 10^{0}$ \\ 
  $250$ & $0.25$ & visGP: Nearest clique & $1.115 \times 10^{-1}$ & 94\% & $1.206 \times 10^{0}$ \\ 
  $250$ & $0.25$ & visGP: Precision-weighted & $1.097 \times 10^{-1}$ & 93\% & $1.165 \times 10^{0}$ \\ 
  $250$ & $0.25$ & visGP: Standard kriging & $1.096 \times 10^{-1}$ & 94\% & $1.197 \times 10^{0}$ \\ 
  $250$ & $0.25$ & Euclidean & $9.344 \times 10^{-1}$ & 81\% & $1.813 \times 10^{0}$ \\ 
  $250$ & $0.25$ & GLGP & $8.290 \times 10^{-1}$ &  &  \\ 
  $250$ & $1$ & BORA-GP & $1.287 \times 10^{0}$ & 95\% & $4.398 \times 10^{0}$ \\ 
  $250$ & $1$ & visGP: Maximum precision & $1.183 \times 10^{0}$ & 94\% & $4.189 \times 10^{0}$ \\ 
  $250$ & $1$ & visGP: Nearest clique & $1.184 \times 10^{0}$ & 94\% & $4.191 \times 10^{0}$ \\ 
  $250$ & $1$ & visGP: Precision-weighted & $1.177 \times 10^{0}$ & 93\% & $4.038 \times 10^{0}$ \\ 
  $250$ & $1$ & visGP: Standard kriging & $1.179 \times 10^{0}$ & 94\% & $4.179 \times 10^{0}$ \\ 
  $250$ & $1$ & Euclidean & $2.015 \times 10^{0}$ & 90\% & $4.532 \times 10^{0}$ \\ 
  $250$ & $1$ & GLGP & $2.047 \times 10^{0}$ &  &  \\ 
  $1200$ & $0.1$ & BORA-GP & $4.703 \times 10^{-2}$ & 96\% & $7.688 \times 10^{-1}$ \\ 
  $1200$ & $0.1$ & visGP: Maximum precision & $3.765 \times 10^{-2}$ & 88\% & $5.265 \times 10^{-1}$ \\ 
  $1200$ & $0.1$ & visGP: Nearest clique & $3.790 \times 10^{-2}$ & 88\% & $5.277 \times 10^{-1}$ \\ 
  $1200$ & $0.1$ & visGP: Precision-weighted & $3.762 \times 10^{-2}$ & 88\% & $5.225 \times 10^{-1}$ \\ 
  $1200$ & $0.1$ & visGP: Standard kriging & $3.760 \times 10^{-2}$ & 88\% & $5.261 \times 10^{-1}$ \\ 
  $1200$ & $0.1$ & Euclidean & $1.037 \times 10^{0}$ & 80\% & $9.068 \times 10^{-1}$ \\ 
  $1200$ & $0.25$ & BORA-GP & $1.190 \times 10^{-1}$ & 96\% & $1.342 \times 10^{0}$ \\ 
  $1200$ & $0.25$ & visGP: Maximum precision & $9.795 \times 10^{-2}$ & 93\% & $1.121 \times 10^{0}$ \\ 
  $1200$ & $0.25$ & visGP: Nearest clique & $9.807 \times 10^{-2}$ & 93\% & $1.122 \times 10^{0}$ \\ 
  $1200$ & $0.25$ & visGP: Precision-weighted & $9.785 \times 10^{-2}$ & 93\% & $1.112 \times 10^{0}$ \\ 
  $1200$ & $0.25$ & visGP: Standard kriging & $9.778 \times 10^{-2}$ & 93\% & $1.121 \times 10^{0}$ \\ 
  $1200$ & $0.25$ & Euclidean & $1.098 \times 10^{0}$ & 82\% & $1.463 \times 10^{0}$ \\ 
  $1200$ & $1$ & BORA-GP & $1.212 \times 10^{0}$ & 95\% & $4.269 \times 10^{0}$ \\ 
  $1200$ & $1$ & visGP: Maximum precision & $1.144 \times 10^{0}$ & 95\% & $4.151 \times 10^{0}$ \\ 
  $1200$ & $1$ & visGP: Nearest clique & $1.145 \times 10^{0}$ & 95\% & $4.152 \times 10^{0}$ \\ 
  $1200$ & $1$ & visGP: Precision-weighted & $1.144 \times 10^{0}$ & 94\% & $4.113 \times 10^{0}$ \\ 
  $1200$ & $1$ & visGP: Standard kriging & $1.143 \times 10^{0}$ & 95\% & $4.149 \times 10^{0}$ \\ 
  $1200$ & $1$ & Euclidean & $2.067 \times 10^{0}$ & 89\% & $4.339 \times 10^{0}$ \\ 
  $10000$ & $0.1$ & BORA-GP & $4.596 \times 10^{-2}$ & 95\% & $6.992 \times 10^{-1}$ \\ 
  $10000$ & $0.1$ & visGP: Maximum precision & $3.787 \times 10^{-2}$ & 84\% & $4.657 \times 10^{-1}$ \\ 
  $10000$ & $0.1$ & visGP: Nearest clique & $3.792 \times 10^{-2}$ & 84\% & $4.659 \times 10^{-1}$ \\ 
  $10000$ & $0.1$ & visGP: Precision-weighted & $3.786 \times 10^{-2}$ & 84\% & $4.645 \times 10^{-1}$ \\ 
  $10000$ & $0.1$ & visGP: Standard kriging & $3.786 \times 10^{-2}$ & 84\% & $4.656 \times 10^{-1}$ \\ 
  $10000$ & $0.1$ & Euclidean & $1.131 \times 10^{0}$ & 83\% & $7.021 \times 10^{-1}$ \\ 
  $10000$ & $0.25$ & BORA-GP & $1.157 \times 10^{-1}$ & 96\% & $1.331 \times 10^{0}$ \\ 
  $10000$ & $0.25$ & visGP: Maximum precision & $9.988 \times 10^{-2}$ & 91\% & $1.070 \times 10^{0}$ \\ 
  $10000$ & $0.25$ & visGP: Nearest clique & $9.993 \times 10^{-2}$ & 91\% & $1.071 \times 10^{0}$ \\ 
  $10000$ & $0.25$ & visGP: Precision-weighted & $9.986 \times 10^{-2}$ & 91\% & $1.068 \times 10^{0}$ \\ 
  $10000$ & $0.25$ & visGP: Standard kriging & $9.983 \times 10^{-2}$ & 91\% & $1.070 \times 10^{0}$ \\ 
  $10000$ & $0.25$ & Euclidean & $1.171 \times 10^{0}$ & 84\% & $1.351 \times 10^{0}$ \\ 
  $10000$ & $1$ & BORA-GP & $1.211 \times 10^{0}$ & 95\% & $4.301 \times 10^{0}$ \\ 
  $10000$ & $1$ & visGP: Maximum precision & $1.191 \times 10^{0}$ & 94\% & $4.099 \times 10^{0}$ \\ 
  $10000$ & $1$ & visGP: Nearest clique & $1.192 \times 10^{0}$ & 94\% & $4.099 \times 10^{0}$ \\ 
  $10000$ & $1$ & visGP: Precision-weighted & $1.191 \times 10^{0}$ & 94\% & $4.086 \times 10^{0}$ \\ 
  $10000$ & $1$ & visGP: Standard kriging & $1.191 \times 10^{0}$ & 94\% & $4.098 \times 10^{0}$ \\ 
  $10000$ & $1$ & Euclidean & $2.030 \times 10^{0}$ & 89\% & $4.342 \times 10^{0}$ \\ 
   \hline
\caption{Simulation results in the fork-shaped domain. Columns give the sample size, standard deviation of the nugget, estimation method, mean square prediction error, confidence/credible interval coverage probability, and mean confidence interval length, respectively.}
\label{tab:sim_fork_res}
\end{longtable}

\newpage

\section{\blue{Plots for Section \ref{web_app_F}}}\label{web_fig_1}

\newpage
\section{Buffered Chesapeake domain}\label{web_fig_2}

\newpage
\blue{\section{Examples of non-convex domains of Proposition \ref{prop:convex_union}}}\label{web_fig_3}
\begin{figure}[ht]
    \centering
    \includegraphics[trim={0 0 300 0}, clip, scale=0.5]{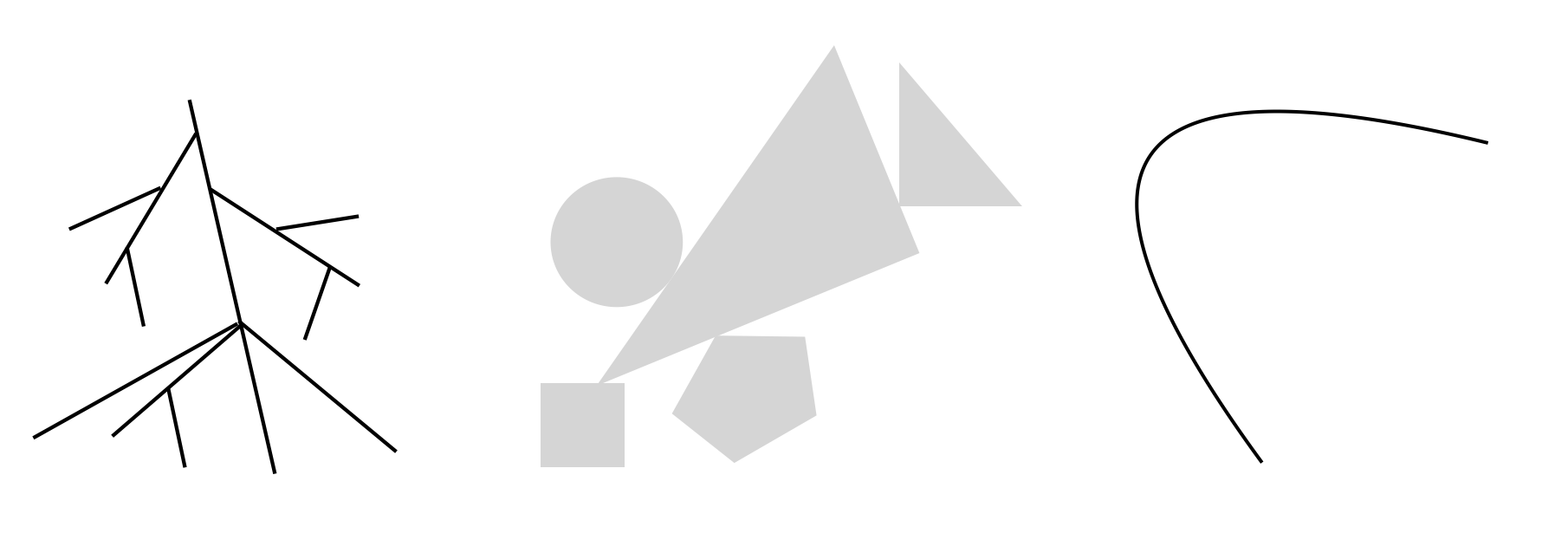}
     \caption{Examples of domains where the covariance function of visGP coincides with an exponential covariance with geodesic distance}\label{fig:union}
\end{figure}

\newpage
\section{\blue{Results in the fork-shaped domain with uniformly random (dense) holdout set}}\label{web_tab_2}
\blue{Note that in most situations, initial parameter estimates for visGP and hyperparameters for BORA-GP came from a BRISC model. In the large sample scenarios below, the parameter estimates from BRISC demonstrated some numeric instability, so we used default values instead.}
\begin{table}[ht]
\centering
\blue{
\begin{tabular}{rrlrrr}
  \hline
n & Nugget.sd & Method & MSE & Coverage & CI.length \\ 
  \hline
$250$ & $0.1$ & BORA GP & $1.436 \times 10^{-2}$ & $9.798 \times 10^{-1}$ & $6.809 \times 10^{-1}$ \\ 
  $250$ & $0.1$ & visGP: Maximum precision & $1.294 \times 10^{-2}$ & $9.481 \times 10^{-1}$ & $4.550 \times 10^{-1}$ \\ 
  $250$ & $0.1$ & GLGP & $1.798 \times 10^{-2}$ &  &  \\ 
  $250$ & $0.25$ & BORA GP & $7.861 \times 10^{-2}$ & $9.611 \times 10^{-1}$ & $1.201 \times 10^{0}$ \\ 
  $250$ & $0.25$ & visGP: Maximum precision & $7.207 \times 10^{-2}$ & $9.490 \times 10^{-1}$ & $1.053 \times 10^{0}$ \\ 
  $250$ & $0.25$ & GLGP & $8.157 \times 10^{-2}$ &  &  \\ 
  $250$ & $1$ & BORA GP & $1.113 \times 10^{0}$ & $9.501 \times 10^{-1}$ & $4.170 \times 10^{0}$ \\ 
  $250$ & $1$ & visGP: Maximum precision & $1.100 \times 10^{0}$ & $9.465 \times 10^{-1}$ & $4.068 \times 10^{0}$ \\ 
  $250$ & $1$ & GLGP & $1.121 \times 10^{0}$ &  &  \\ 
  $1200$ & $0.1$ & BORA GP & $1.226 \times 10^{-2}$ & $9.594 \times 10^{-1}$ & $4.642 \times 10^{-1}$ \\ 
  $1200$ & $0.1$ & visGP: Maximum precision & $1.129 \times 10^{-2}$ & $9.486 \times 10^{-1}$ & $4.148 \times 10^{-1}$ \\ 
  $1200$ & $0.25$ & BORA GP & $7.035 \times 10^{-2}$ & $9.541 \times 10^{-1}$ & $1.063 \times 10^{0}$ \\ 
  $1200$ & $0.25$ & visGP: Maximum precision & $6.889 \times 10^{-2}$ & $9.482 \times 10^{-1}$ & $1.022 \times 10^{0}$ \\ 
  $1200$ & $1$ & BORA GP & $1.096 \times 10^{0}$ & $9.478 \times 10^{-1}$ & $4.094 \times 10^{0}$ \\ 
  $1200$ & $1$ & visGP: Maximum precision & $1.099 \times 10^{0}$ & $9.469 \times 10^{-1}$ & $4.081 \times 10^{0}$ \\ 
  $10000$ & $0.1$ & BORA GP & $1.115 \times 10^{-2}$ & $9.534 \times 10^{-1}$ & $4.204 \times 10^{-1}$ \\ 
  $10000$ & $0.1$ & visGP: Maximum precision & $1.109 \times 10^{-2}$ & $9.467 \times 10^{-1}$ & $4.081 \times 10^{-1}$ \\ 
  $10000$ & $0.25$ & BORA GP & $6.910 \times 10^{-2}$ & $9.482 \times 10^{-1}$ & $1.024 \times 10^{0}$ \\ 
  $10000$ & $0.25$ & visGP: Maximum precision & $6.968 \times 10^{-2}$ & $9.456 \times 10^{-1}$ & $1.019 \times 10^{0}$ \\ 
  $10000$ & $1$ & BORA GP & $1.089 \times 10^{0}$ & $9.491 \times 10^{-1}$ & $4.083 \times 10^{0}$ \\ 
  $10000$ & $1$ & visGP: Maximum precision & $1.104 \times 10^{0}$ & $9.468 \times 10^{-1}$ & $4.081 \times 10^{0}$ \\ 
   \hline
\end{tabular}}
\end{table}

\bibliographystyle{plainnat}
\bibliography{graph_latex_arXiv/references}

\end{document}